\documentclass[sigconf]{acmart}
\usepackage{graphicx}

\usepackage{amsmath}
\usepackage{graphicx}
\usepackage{color}
\usepackage{amsfonts}
\usepackage{verbatimbox}

\DeclareGraphicsExtensions{.pdf, .png}

\usepackage{enumitem} %

\usepackage{caption} %
\usepackage{subcaption} %

\usepackage{cellspace}
\newcommand\cincludegraphics[2][]{\raisebox{-0.3\height}{\includegraphics[#1]{#2}}}

\usepackage{wrapfig} %

\newcommand{\dset}{\mathcal{X}}
\newcommand{\dpoint}{\boldsymbol{x}}
\newcommand{\dpointlat}{\textit{lat}}
\newcommand{\dpointlon}{\textit{lon}}
\newcommand{\dpointtime}{\textit{t}}
\newcommand{\dpointsigma}{\sigma}
\newcommand{\dpointpred}{\hat{\boldsymbol{x}}}
\newcommand{\traj}{S}
\newcommand{\trajdegraded}{Z}
\newcommand{\dpointdegraded}{z}
\newcommand{\sigmadegrade}{\sigma_z}
\newcommand{\addednoise}{\eta}
\newcommand{\truncationratio}{\alpha_t}
\newcommand{\subratio}{\alpha_s}

\newcommand{\priorinfo}{\mathbf{\Omega}}
\newcommand{\probdist}{P}
\newcommand{\trajtimegap}{\tau}
\newcommand{\corrcoeff}{\rho}
\newcommand{\duration}{D}
\newcommand{\traveldist}{L}

\newcommand{\entropy}{H}
\newcommand{\entropyspatial}{\entropy_S}
\newcommand{\entropytemporal}{\entropy_T}
\newcommand{\ig}{\textit{IG}}
\newcommand{\diffentropy}{h}
\newcommand{\igperiod}{T}

\newcommand{\gppredset}{\mathcal{T}}

\copyrightyear{2021} 
\acmYear{2021} 
\setcopyright{acmlicensed}
\acmConference[SIGSPATIAL '21]{29th International Conference on Advances in Geographic Information Systems}{November 2--5, 2021}{Beijing, China}
\acmBooktitle{29th International Conference on Advances in Geographic Information Systems (SIGSPATIAL '21), November 2--5, 2021, Beijing, China}
\acmPrice{15.00}
\acmDOI{10.1145/3474717.3483912}
\acmISBN{978-1-4503-8664-7/21/11}

\begin{document}

\title{Quantifying Intrinsic Value of Information of Trajectories}

\author{Kien Nguyen}
\email{kien.nguyen@usc.edu}
\affiliation{%
  \institution{University of Southern California}
  \city{Los Angeles}
  \state{California}
  \country{USA}
}

\author{John Krumm}
\email{jckrumm@microsoft.com}
\affiliation{%
  \institution{Microsoft Research}
  \city{Redmond}
  \state{Washington}
  \country{USA}
}

\author{Cyrus Shahabi}
\email{shahabi@usc.edu}
\affiliation{%
  \institution{University of Southern California}
  \city{Los Angeles}
  \state{California}
  \country{USA}
}

\renewcommand{\shortauthors}{Nguyen, et al.}



\begin{abstract}
A trajectory, defined as a sequence of location measurements, contains valuable information about movements of an individual. Its value of information (VOI) may change depending on the specific application. However, in a variety of applications, knowing the intrinsic VOI of a trajectory is important to guide other subsequent tasks or decisions. This work aims to find a principled framework to quantify the intrinsic VOI of trajectories from the owner's perspective. This is a challenging problem because an appropriate framework needs to take into account various characteristics of the trajectory, prior knowledge, and different types of trajectory degradation. We propose a framework based on information gain (IG) as a principled approach to solve this problem. Our IG framework transforms a trajectory with discrete-time measurements to a canonical representation, i.e., continuous in time with continuous mean and variance estimates, and then quantifies the reduction of uncertainty about the locations of the owner over a period of time as the VOI of the trajectory. Qualitative and extensive quantitative evaluation show that the IG framework is capable of effectively capturing important characteristics contributing to the VOI of trajectories. 
\end{abstract}

\maketitle

\section{Introduction}
\label{sec:intro}
The availability of mobile devices with location-tracking capability has enabled individuals to generate a great amount of location data from various types of signals, e.g., GPS, Wi-Fi, or cell service. A trajectory, which is a sequence of location measurements, is an important type of location data that contains valuable information about the owner's locations and movements. For example, a trajectory may indicate the owner's moving behavior~\cite{yue2019detect}, e.g., a trip to home, office, or favorite shops and time spent there; it may help identify irregularities such as vacation days or house moving; it is important for traffic, speed, and route inference~\cite{zheng2015trajectory}. 

Each trajectory gives information of some kind. The value of the information (VOI) varies depending on the use and user of the data. For example, the same trajectory may have one value for the owner and another for an enterprise, and that same trajectory may also have different values for different enterprises, such as a public or private, especially for an ad-targeting company. 

However, a trajectory also contains an intrinsic, formulaic VOI, since it contains quantified locations of an individual over space and time. For example, it is reasonable to assume that a home-to-office trajectory with 1000 measurements from beginning to end of the trip tends to have more information than another home-to-office trajectory with only 2 measurements at home and office. Our goal is to \emph{find a principled approach to quantify the intrinsic VOI of trajectories from the owner's perspective}. To the best of our knowledge, this is the first attempt at such a quantification. 

Given a trajectory, there are natural questions about selling it~\cite{kanza2015geomarketplace}, sharing it, using it to train machine learning, storing it, or examining it more closely. Quantifying its VOI helps measure how valuable, revealing, informative~\cite{varshneya2017human}, distinct, and surprising~\cite{zheng2015trajectory} it is. A trajectory's VOI can be a critical piece of metadata that indicates its intrinsic value for a variety of tasks, both stand-alone and as part of a collection.

This problem presents several interesting challenges:
\begin{itemize}[nosep]
    \item First, a trajectory has many characteristics contributing to its VOI, and effectively capturing their complex relationships is non-trivial. Examples of these characteristics are the number of measurements, temporal duration, and how measurements distribute spatially and temporally. %
    \item Second, a trajectory can be degraded for different purposes, e.g., measurements can be perturbed by adding random noise or completely removed to enhance the owner's privacy before releasing or selling. It is challenging to capture the effect of different types of degradation on the VOI of the trajectory.
    \item Third, the owner may assume different prior knowledge about a trajectory, which may change its VOI from the owner's perspective. For example, if the owner has never released any trajectory data before, meaning their trajectory would give more information about their locations than if they already released a perturbed version of the trajectory. 
\end{itemize}   

Some straightforward methods, e.g., using a trajectory's characteristics such as size or duration, fail to capture other characteristics or prior knowledge. Previous work based on Spatial Privacy Pricing ~\cite{nguyen2020spatialprivacypricing} can be adapted to sum the values of degraded measurements of the trajectory. However, it also fails to capture some characteristics (e.g, how measurements distribute over space) and prior knowledge, because it ignores the mutual information of the constituent points. The notion of correctness~\cite{shokri2011quantifying} also appears promising. For this method, based on a degraded version $\trajdegraded$ of a trajectory $\traj$ and prior knowledge, a probabilistic prediction is made for each measurement. Then the correctness of the prediction, indicating how close it is to the actual measurement, is aggregated to derive the VOI of $\trajdegraded$. However, this method is not applicable when $\trajdegraded = \traj$, because correctness is not available when there is no actual measurement with which to compare the prediction.  

The core idea behind our approach is that instead of computing the VOI of a trajectory from its discrete location measurements over time, we view VOI as how much the trajectory data helps reconstruct locations of the owner. More specifically, the VOI corresponds to how much the trajectory data helps reduce the uncertainty of estimating its owner's locations in continuous time. We use information gain (IG) to quantify that reduction. Even though IG is a known measure~\cite{quinlan1986induction}, it is not obvious how it should be used to compare discrete-time trajectories with widely different characteristics: long and short, dense and sparse, clean and noisy. Our innovation here is that we transform each trajectory to a canonical representation, i.e., continuous in time with continuous estimates of mean and variance, which is necessary for comparing widely different trajectories and comparing new data against prior data. IG can only then be applied  after all those steps are taken. We realize this transformation by employing a reconstruction method that can produce continuous-time probabilistic predictions along the trajectory.

Consequently, we propose an IG framework as a principled way to quantify the intrinsic VOI of a trajectory. The main idea is to quantify the reduction of uncertainty about the owner's continuous-time locations, comparing a new trajectory to a previously released degraded version or to prior information. Thus the IG framework utilizes a reconstruction method to produce continuous probabilistic predictions over time, and then it calculates the reduction of uncertainty from those predictions compared to a reconstruction from prior knowledge. The uncertainty is measured by (differential) entropy. Our IG framework accepts any reasonable probabilistic reconstruction method. Gaussian process is used as the reconstruction method in this paper due to its popularity and flexibility, but is not necessarily the only choice.

Specifically, our contributions are:
\begin{itemize}[nosep]
    \item Propose the problem of quantifying intrinsic VOI of a trajectory from the owner's perspective
    \item Define characteristics that should be captured by an appropriate quantification method
    \item Show how alternate methods fail, even when they appear reasonable initially
    \item Introduce a method to transform each trajectory to a canonical representation, allowing us to examine the VOI with a continuous location reconstruction and to compare trajectories with widely different characteristics
    \item Develop an IG framework over transformed trajectories as a principled approach capable of effectively capturing various trajectory characteristics, prior knowledge, and degradation
    \item Evaluate the proposed framework both qualitatively and quantitatively with extensive experiments on a large, real-world trajectory dataset
\end{itemize}
By providing a standard, comprehensive method for assessing the VOI of a trajectory, we enable a deeper understanding of trajectories and their utility for a wide variety of purposes.

\section{Problem Setting}
\label{sec:problem_setting}

This section introduces the problem of quantifying the intrinsic VOI of a trajectory from the owner's perspective. We focus on the most basic form of a trajectory, which is a sequence of location measurements. Therefore, incorporating other information, such as census data or points of interest, is beyond the scope of this work.

The owner can be anyone having access and right to use the data, e.g., one whose phone recorded this trajectory, or a data collector who aggregates location data from individuals. Without loss of generality and to ease the discussion, from here on, we assume the owner is the individual whose device recorded the trajectory.

A trajectory $\traj$ is a sequence of (potentially noisy) location measurements $\traj = \{ \dpoint_1, \dpoint_2, \dots, \dpoint_{|\traj|} \}$. Each measurement or data point $\dpoint$ (bold symbol) is a tuple $\dpoint =\, <\dpointlon, \dpointlat, \dpointtime, \dpointsigma>$ where $\dpoint.\dpointlon$ and $\dpoint.\dpointlat$ are the longitude and latitude, $\dpoint.\dpointtime$ is the timestamp, and $\dpoint.\dpointsigma$ is the accuracy or uncertainty of the measurement. It is reasonable to assume Gaussian noise for location measurements~\cite{diggelen2007system} such as GPS points, so  $\dpoint.\dpointsigma$ can be considered as the standard deviation of independent Gaussian noise of longitude and latitude. Measurements in $\traj$ are ordered by their timestamps, i.e, $\forall \dpoint_i, \dpoint_j \in \traj, 1 \leq i \leq j \leq |\traj|, \dpoint_i.\dpointtime \leq \dpoint_j.\dpointtime$.

The owner can assume the potential recipient has some prior knowledge~$\priorinfo$ about the location $\dpoint_i$, e.g., by gathering public data or from some data already released from the owner before. This prior knowledge is represented as a prior distribution $\probdist(\dpoint_i | \priorinfo)$ and should be respected by the proposed methods. Intuitively, prior knowledge about a point would tend to reduce the VOI of the point.

In addition to releasing raw trajectories, owners can also offer their data at different quality levels, potentially enhancing their privacy while reducing the trajectory's VOI, e.g, a trajectory can be degraded by adding noise or subsampling. More details about degradation are discussed in Section~\ref{sec:trajectory_degradation}. The methods to quantify the VOI also need to reflect this quality degradation.

Our goal is to find a principled approach to quantify the VOI of a trajectory from the owner's perspective. It should capture different characteristics of the raw trajectory, the prior knowledge, and the effect of various degradation processes. 
Next, we described the characteristics that an appropriate method should aim to capture. This is an extensive but not exhaustive list of desirable characteristics, thus we expect future work will study more characteristics. We will then illustrate how several baseline methods fail to capture these characteristics and how our new proposed framework succeeds.

\subsection{Trajectory Characteristics}
From the raw trajectory $\traj$, certain characteristics can be derived that should contribute to its VOI. These includes \emph{size}, \emph{duration}, \emph{spatial distribution}, \emph{temporal distribution}, and \emph{measurement uncertainty}. 

\textbf{\textit{Size.}} The size $|\traj|$ is the number of measurements of $\traj$. A larger size often means there is more data to learn about the owners' locations and movements. Therefore, it is reasonable to assume that a trajectory with a larger size tends to have more information.

\textbf{\textit{Duration.}} The duration of a trajectory $\traj$ is the time difference between the first and last measurements of $\traj$, i.e, $\duration(\traj) = \dpoint_{|\traj|}.t - \dpoint_1.t$. Similar to the size, a trajectory with a longer duration often means it may have more information about the owner's locations than one with a shorter duration.

\textbf{\textit{Spatial distribution.}} The spatial distribution of $\traj$ indicates how its measurements are distributed over space. For example, one trajectory may have all measurements at one place, while another trajectory may have measurements at multiple places (e.g, solely at home vs. at home, then at a coffee shop, then at the office). A trajectory visiting more places tends to give more information about locations of the owner. The spatial distribution of $\traj$ can be measured using \emph{spatial entropy} $\entropyspatial(\traj)$ over a grid of cells, say $10m \times 10m$, covering the area of interest. $\entropyspatial(\traj)$ is calculated by creating a histogram of number of measurements of $\traj$ belonging to each cell, converting this histogram to probabilities, and then computing the Shannon entropy for the probabilities. 

\textbf{\textit{Temporal distribution.}} Similarly, the temporal distribution of a trajectory $\traj$ indicates how its measurements are distributed over time. For example, one trajectory may have many measurements in the first few minutes and then a long gap before the next one. Another trajectory may be distributed more uniformly over time, e.g., one measurement every 30 seconds. Intuitively, when measurements distribute more evenly over a longer period of time, they tend to give more information. The temporal distribution of $\traj$ can be measured using the \emph{temporal entropy} $\entropytemporal(\traj)$ calculated during $[\dpoint_1.\dpointtime, \dpoint_{|\traj|}.\dpointtime]$ by computing a histogram of the number of measurements of $\traj$ belonging to each temporal bin, say every 1 minute, from $\dpoint_1.\dpointtime$ to $\dpoint_{|\traj|}.\dpointtime$, converting the histogram to probabilities, and then computing the Shannon entropy for the probabilities.

\textbf{\textit{Measurement uncertainty.}} Measurements may have their own uncertainty depending on how they were taken, e.g, an uncertainty of several hundred meters with cell towers~\cite{chen2006practical} or 1 to 5 meters for GPS-enabled smartphones~\cite{gpsaccuracy}. Since Gaussian noise is a reasonable assumption for GPS~\cite{diggelen2007system}, measurement uncertainty is defined as the standard deviation $\dpointsigma$ of independent Gaussian noise of longitude and latitude. In general, less accurate measurements (i.e, larger $\dpointsigma$) tend to give less information about locations of the owner.

\subsection{Prior Knowledge}
As mentioned, the owner can assume there exists some available prior knowledge about the location $\dpoint_i$, e.g., from public data or from previous releases from this owner. This knowledge is expressed as a prior distribution $\probdist(\dpoint_i | \priorinfo)$ for each timestamp of interest. In general, better prior knowledge often means the trajectory gives less information compared to the case with poorer prior knowledge.

\subsection{Trajectory Degradation}
\label{sec:trajectory_degradation}
The owner can offer their trajectories at lower quality. The process of lowering the quality of a trajectory is called \emph{degradation}. The reason for the owner to produce degraded trajectories is that a recipient can still benefit from data at a certain quality depending on their specific applications~\cite{nguyen2020spatialprivacypricing}. For example, estimating a neighborhood-level origin-destination matrix may not need extremely accurate measurements, thus the recipient with this application can potentially spend less to purchase lower quality trajectories. In general, a lower quality trajectory would give less information than a higher quality one. A degraded version of a trajectory $\traj$ is denoted as $\trajdegraded$. 

In this work, we consider three types of degradation: \emph{perturbation}, \emph{truncation}, and \emph{subsampling}. Other types of degradations and other variations of these degradations are beyond the scope of this paper and considered as part of future work.

\textbf{\textit{Perturbation.}} The perturbation process degrades a trajectory $\traj$ by adding independent random Gaussian noise with standard deviation $\sigmadegrade$ to each $\dpoint_i \in \traj$. Formally, a perturbed trajectory $\trajdegraded$ of $\traj$ consists of measurements $\dpointdegraded_i$ s.t.
\begin{align}
    \dpointdegraded_i.\dpointlon &= \dpoint_i.\dpointlon + \addednoise_1 \\ \dpointdegraded_i.\dpointlat &= \dpoint_i.\dpointlat + \addednoise_2 \\ \dpointdegraded_i.\dpointtime &= \dpoint_i.\dpointtime \\ 
    \dpointdegraded_i.\dpointsigma &= \sqrt{(\dpoint_i.\dpointsigma)^2 + \sigmadegrade^2} \\
    \addednoise_1, \addednoise_2 &\sim \mathcal{N}(0, \sigmadegrade^2)
\end{align}
Noise magnitude $\dpointdegraded_i.\dpointsigma$ is called the \emph{total noise} of $\trajdegraded$.

\textbf{\textit{Truncation.}} The truncation process degrades a trajectory $\traj$ by truncating $\traj$ and only keeping a fraction $\truncationratio \in [0, 1]$\ of the first measurements of $\traj$. For example, if $|\traj| = 100$, an $\truncationratio = 0.2$ means that the temporally first $20\%$ of measurements are kept, which are $\{\dpoint_1, \dots, \dpoint_{20}\}$. The fraction $\truncationratio$ is called the \emph{truncation ratio}. The retained measurements are kept in their raw form. Also, there is at least one measurement retained. 

\textbf{\textit{Subsampling.}} Similarly, the subsampling process degrades a trajectory $\traj$ by uniformly subsampling $\traj$ with probablity $\subratio \in [0, 1]$, called the \emph{subsampling ratio}. More specifically, a measurement $\dpoint_i\in \traj$ is retained with probability $\subratio$; otherwise, $\dpoint_i$ is discarded. Hence, in expectation, a fraction $\subratio$ of measurements of $\traj$ are retained. The retained measurements are also kept in their raw form and at least one measurement is retained.

\section{Baselines}
\label{sec:baselines}
\begin{table*}[t]
\begin{tabular}{l||Sc|Sc|Sc|Sc|Sc|Sc|Sc}
                          & \textbf{Size} & \textbf{Duration} & \textbf{\begin{tabular}[c]{@{}l@{}}Spatial\\ Distribution\end{tabular}} & \textbf{\begin{tabular}[c]{@{}l@{}}Temporal\\ Distribution\end{tabular}} & \textbf{\begin{tabular}[c]{@{}l@{}}Measurement\\ Uncertainty\end{tabular}} & \textbf{\begin{tabular}[c]{@{}l@{}}Prior\\ Knowledge\end{tabular}} & \textbf{Degradation} \\ \hline \hline
\textbf{Fixed Value}      & 
\cincludegraphics[height=0.22cm]{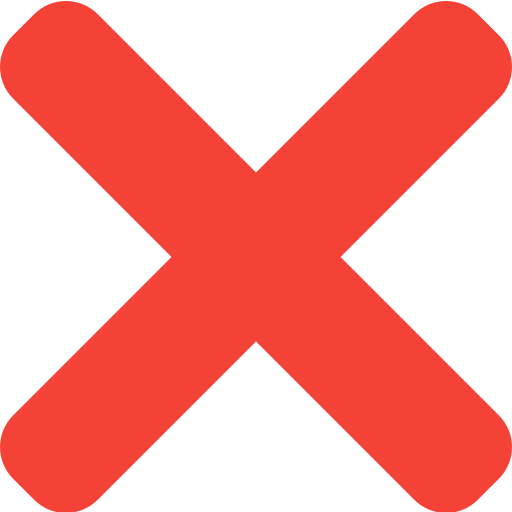}  &     \cincludegraphics[height=0.22cm]{figures/cross.png}  &  \cincludegraphics[height=0.22cm]{figures/cross.png}  & \cincludegraphics[height=0.22cm]{figures/cross.png}  &                                   \cincludegraphics[height=0.22cm]{figures/cross.png}  & \cincludegraphics[height=0.22cm]{figures/cross.png}  & \cincludegraphics[height=0.22cm]{figures/cross.png}                        \\ \hline
\textbf{Size-based}      & 
\cincludegraphics[height=0.22cm]{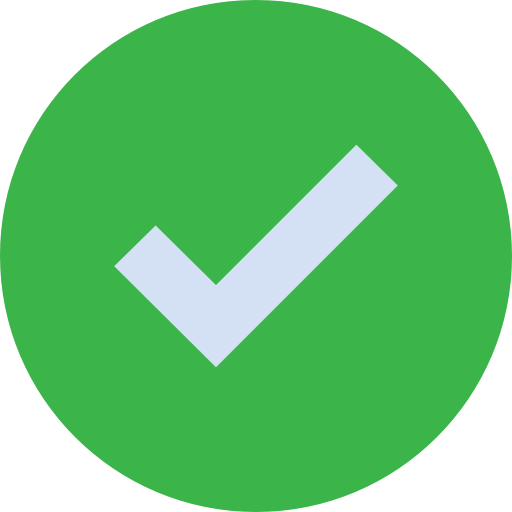}  &     \cincludegraphics[height=0.22cm]{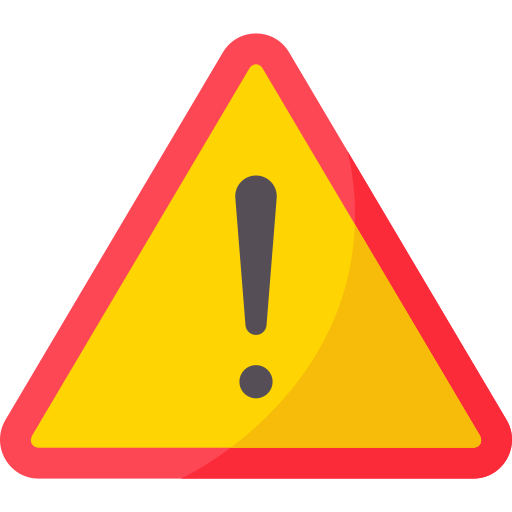}  &  \cincludegraphics[height=0.22cm]{figures/cross.png}  & \cincludegraphics[height=0.22cm]{figures/cross.png}  &                                   \cincludegraphics[height=0.22cm]{figures/cross.png}  & \cincludegraphics[height=0.22cm]{figures/cross.png}  & \cincludegraphics[height=0.22cm]{figures/alert.png}                        \\ \hline
\textbf{Duration-based}      & 
\cincludegraphics[height=0.22cm]{figures/alert.png}  &     \cincludegraphics[height=0.22cm]{figures/check.png}  &  \cincludegraphics[height=0.22cm]{figures/cross.png}  & \cincludegraphics[height=0.22cm]{figures/cross.png}  &                                   \cincludegraphics[height=0.22cm]{figures/cross.png}  & \cincludegraphics[height=0.22cm]{figures/cross.png}  & \cincludegraphics[height=0.22cm]{figures/alert.png}                        \\ \hline
\textbf{Travel Distance}      & 
\cincludegraphics[height=0.22cm]{figures/alert.png}  &     \cincludegraphics[height=0.22cm]{figures/alert.png}  &  \cincludegraphics[height=0.22cm]{figures/alert.png}  & \cincludegraphics[height=0.22cm]{figures/alert.png}  &                                   \cincludegraphics[height=0.22cm]{figures/cross.png}  & \cincludegraphics[height=0.22cm]{figures/cross.png}  & \cincludegraphics[height=0.22cm]{figures/alert.png}                        \\ \hline
\textbf{Entropy-based}      & 
\cincludegraphics[height=0.22cm]{figures/alert.png}  &     \cincludegraphics[height=0.22cm]{figures/alert.png}  &  {}\cincludegraphics[height=0.22cm]{figures/check.png}*  & {}\cincludegraphics[height=0.22cm]{figures/check.png}*  &                                   \cincludegraphics[height=0.22cm]{figures/cross.png}  & \cincludegraphics[height=0.22cm]{figures/cross.png}  & \cincludegraphics[height=0.22cm]{figures/alert.png}                        \\ \hline
\textbf{SPP-based~\cite{nguyen2020spatialprivacypricing}}      & \cincludegraphics[height=0.22cm]{figures/check.png}  &     \cincludegraphics[height=0.22cm]{figures/alert.png}  &  \cincludegraphics[height=0.22cm]{figures/cross.png}  & \cincludegraphics[height=0.22cm]{figures/cross.png}  &                                   \cincludegraphics[height=0.22cm]{figures/check.png}  & \cincludegraphics[height=0.22cm]{figures/cross.png}  & \cincludegraphics[height=0.22cm]{figures/check.png}                        \\ \hline
\textbf{Correctness-based~\cite{shokri2011quantifying}}      & \cincludegraphics[height=0.22cm]{figures/alert.png}  &     \cincludegraphics[height=0.22cm]{figures/alert.png}  &  \cincludegraphics[height=0.22cm]{figures/alert.png}  & \cincludegraphics[height=0.22cm]{figures/alert.png}  &                                   \cincludegraphics[height=0.22cm]{figures/alert.png}  & { }\cincludegraphics[height=0.22cm]{figures/check.png}*  & { }\cincludegraphics[height=0.22cm]{figures/check.png}*                       \\ \hline
\textbf{Information Gain}      & 
\cincludegraphics[height=0.22cm]{figures/check.png}  &     \cincludegraphics[height=0.22cm]{figures/check.png}  &  \cincludegraphics[height=0.22cm]{figures/alert.png}  & \cincludegraphics[height=0.22cm]{figures/check.png}  &                                   \cincludegraphics[height=0.22cm]{figures/check.png}  & \cincludegraphics[height=0.22cm]{figures/check.png}  & \cincludegraphics[height=0.22cm]{figures/check.png}                 
\end{tabular}
\caption{Potential methods to quantify the intrinsic VOI of a trajectory and the characteristics they can or cannot capture. Our proposed method based on information gain can capture almost all of the desirable characteristics.}
\label{tbl:methods_characteristics}
\end{table*}

This section discusses  baseline techniques to quantify the VOI of a trajectory. Each baseline is described along with which desirable characteristics from Section~\ref{sec:problem_setting} they can capture. Table~\ref{tbl:methods_characteristics} summarizes which characteristics the baselines and our proposed method can faithfully represent. Our proposed method, based on information gain and shown in the last row of Table~\ref{tbl:methods_characteristics}, is described in detail in Section~\ref{sec:ig_based}. A cross (\includegraphics[height=\fontcharht\font`\B]{figures/cross.png})/exclamation mark (\includegraphics[height=\fontcharht\font`\B]{figures/alert.png})/checkmark (\includegraphics[height=\fontcharht\font`\B]{figures/check.png}) indicates that the method in that row cannot/can partially/can fully capture the characteristic in that column, respectively. 

The following analysis shows how the baseline methods, while appearing initially reasonable, fail to represent some important characteristics of the VOI. For each characteristic, a method is first evaluated qualitatively. In some cases, it is obvious that a method can or cannot capture a characteristic, e.g., a size-based method can capture the size but not prior knowledge. In other cases, it may not be as clear. In those cases, the method is said to be capable of capturing the characteristic if there is a strong correlation between the output of the method and the characteristic. The correlation is examined using Spearman's rank correlation coefficient $\corrcoeff$, which quantifies strictly monotonic relationships between two variables and is relatively robust against outliers~\cite{schober2018correlation}. An absolute magnitude $|\corrcoeff|$ in $[0.4, 0.7)$ and $[0.7, 1]$ indicates moderate and strong correlation, respectively. The cutoff points are based on previous work~\cite{schober2018correlation}. The values of $\corrcoeff$ are calculated from a large real-world trajectory dataset described in Section~\ref{sec:datasets}.

\subsection{Fixed Value Method} 
One potential method is to set the same value for all trajectories. While this method is straightforward and, in fact, was used in some surveys about values of location data from a seller's perspective~\cite{cvrcek2006study, staiano2014money}, this method ignores the fact that each trajectory may contain a different VOI. For example, a 40-mile long commute from home to work can be very different from a short trip to a nearby store. Hence, this method does not capture any of the desirable characteristics.

\subsection{Size-based Method} 
\label{sec:baseline_size}
The size-based method uses the size $|\traj|$ of a trajectory $\traj$ to quantify the VOI of $\traj$. It is reasonable to say that a trajectory with more measurements tends to have more information. While this method may distinguish trajectories with many or few measurements, the size alone would fail to fully capture the VOI of a trajectory, because the size $|\traj|$ depends heavily on the sampling rate and the duration. For example, the same trip from home to office if sampled every 1 second would have a size 5 times larger than if sampled every 5 seconds, while having roughly similar information about the locations of the person along the trip. 

Consequently, the size-based method successfully captures the size characteristic of a trajectory, but not other aspects. For the duration characteristic, because it only depends on the first and last measurements of the trajectory, the size cannot fully capture it, e.g., a trajectory with two measurements can be arbitrarily short or long. However, the size can partially capture the duration if the sampling rate is relatively similar among trajectories. The correlation coefficient $\corrcoeff$ between size and duration of trajectories in our dataset is $0.86$, which indicates a strong correlation between them.

The size $|\traj|$ also does not capture the spatial nor temporal distribution of $\traj$, because it does not indicate how the measurements are distributed. The same number of measurements can happen at nearly the same place/time if the sampling rate is high, or at different places/times if the sampling rate is low. It is also clear that $|\traj|$ does not indicate measurement uncertainty nor prior knowledge. In fact, any method using solely trajectory characteristics would fail to capture prior knowledge because prior knowledge is not considered in that method. The size can capture some degradation such as truncation, but not other degradation such as perturbation. %

\subsection{Duration-based Method}
Another method is to use the duration $\duration(\traj)$ of $\traj$, i.e, $\duration(\traj) = \dpoint_{|\traj|}.t - \dpoint_1.t$, to quantify its VOI, as it is reasonable to assume that a temporally longer trajectory tends to have more information. However, using only duration would fail to fully capture the VOI since it ignores all information between the start and end points. 

Similar to the size-based method explained before, this method can fully capture the duration but can only partially capture the size. The duration also does not represent how the measurements distribute over space and time, and is unable to take into account measurement uncertainty and prior knowledge. Duration can capture the effect of truncation but not that of perturbation nor subsampling, thus, only partially capturing the effects of degradation.

\subsection{Travel Distance Method} 
This method computes the travel distance $\traveldist(\traj)$ of $\traj$ by summing the distances between each pair of consecutive measurements. 
It is reasonable to assume that a trajectory with longer distance tends to have more information than a shorter one. 

With the same mode of transportation (e.g., with bikes or cars), travel distance likely reflects the duration. Thus, travel distance can partially capture duration and the size of $\traj$. With a longer distance, the trajectory tends to go through more places and time period. In our dataset, a strong correlation $0.89$ and $0.78$ between the travel distance and spatial and temporal entropy indicate that travel distance can partially capture the spatial and temporal distribution.

However, using travel distance would fail take into account measurement uncertainty or any prior knowledge. It can capture some degradation, e.g., truncation, but not others, thus, only partially capturing the effect of degradation.

\subsection{Entropy-based Method} 
The spatial and temporal entropy $\entropyspatial(\traj)$ or $\entropytemporal(\traj)$ of $\traj$, computed from the Shannon entropy of the probabilities converted from the histogram of number of measurements of $\traj$ belonging to each grid cell or temporal bin, can be good candidates to quantify the VOI given its extensive applications in information theory. A higher spatial/temporal entropy likely indicates that the trajectory gives more information about a person's locations over space/time. 

Since $\entropyspatial(\traj)$ or $\entropytemporal(\traj)$ are used to measure the spatial and temporal distribution of $\traj$, respectively, a method combining both $\entropyspatial(\traj)$ or $\entropytemporal(\traj)$ can capture these two characteristics. However, it is unclear how they should be combined, which is why there are asterisks for the entropy-based row in Table~\ref{tbl:methods_characteristics}. 

While $\entropyspatial(\traj)$ and $\entropytemporal(\traj)$ do not fully capture the size and duration (e.g., a trajectory with only two points but far apart from each other can have a long duration but low entropy values), when $\traj$ has a relatively high sampling rate, $\entropytemporal(\traj)$ would be highly correlated with the size and duration. In our dataset, the correlation coefficients between $\entropytemporal(\traj)$ and size and duration are $0.89$ and $0.97$, respectively, which indicate a very strong correlation. 

However, since entropy reflects the uncertainty, when the measurements have higher uncertainty (i.e., larger $\dpoint.\dpointsigma$), both $\entropyspatial(\traj)$ and $\entropytemporal(\traj)$ tend to increase. This is opposite of what one might expect, because a more uncertain measurement means less information. Thus, entropy does not capture measurement uncertainty nor degradation such as perturbation. 
Computing $\entropyspatial(\traj)$ and $\entropytemporal(\traj)$ also ignores all prior knowledge. Another issue is that the entropy can change significantly when arbitrary parameters for computing entropy (i.e., size of grid cells or length of time bins) change.

\subsection{Spatial Privacy Pricing-based Method}
This method is based on the previous work on Spatial Privacy Pricing~\cite{nguyen2020spatialprivacypricing} (SPP) where each measurement has the same value defined by the owner, and the value can be reduced when the measurement is perturbed by noise. The VOI of the trajectory is then calculated by summing up the values of each individual measurement. 

With the ability to change the value based on the noise in the measurements, this method can capture the measurement uncertainty and represent degradation. Summing the individual values means this method is similar to the size-based method, but is also sensitive to the perturbation parameters.  Therefore, this method has similar characteristics as the size-based method, which means fully capturing size, partially capturing duration, and unable to capture spatial and temporal distributions and prior knowledge.

\subsection{Correctness-based Method}
\label{sec:correctness}
This method is based on the \emph{correctness} of reconstructing the raw measurements from available data~\cite{shokri2011quantifying}. Roughly speaking, when calculating the VOI of a degraded version $\trajdegraded$ of $\traj$, the owner can assume the recipient is attempting to reconstruct each measurement $\dpoint_i \in \traj$ based on $\trajdegraded$. The correctness is the expected error between the actual points $\dpoint_i$ and the probabilistically reconstructed points $\dpointpred_i$. If the expected error is lower, then it would be reasonable to assume a higher VOI of $\trajdegraded$.

While this concept was proposed for a different problem setting, some techniques can be used to adapt it to our problem. In fact, our proposed framework, discussed later, also has a reconstruction step that can capture prior knowledge and degradation. Thus a correctness-based method can capture the prior knowledge and degradation. The asterisks in the correctness-based row in Table~\ref{tbl:methods_characteristics} indicate that a correctness-based method needs some modifications for the problem setting and techniques to be fit for our problem.

The main drawback of using correctness is that it requires some ground-truth measurements of $\traj$ being available to evaluate the correctness of the reconstructed trajectory made from the degraded version $\trajdegraded$ and prior knowledge. So, it cannot measure the VOI of the full, raw trajectory $\traj$, because there is no ground-truth measurement exists to evaluate correctness in that case. Consequently, since the full, raw trajectory cannot be fully captured, which means there are often some characteristics not fully captured (e.g., subsampling changes the size and truncation changes the duration), this method can only partially capture characteristics of a trajectory. 

Another issue is that this method relies on the correctness of discrete-time predictions, and it is unclear how the correctness of each prediction should be aggregated to obtain the correctness of the whole trajectory. For example, the correctness derived from an $80\%$ subsampled version is evaluated on each measurement of the remaining $20\%$ data, while the correctness derived from a $60\%$ subsampled version is evaluated on the remaining $40\%$. It is unclear how the correctness should be modified to reasonably quantify both, and how the correctness of each prediction should be aggregated.

\subsection{Other Potential Quantities}
There are other potential quantities contributing to the VOI of a trajectory. However, these are not considered as baselines because they are either orthogonal (i.e., they can be used in conjunction with other methods) or complicated (i.e., finding these quantities requires techniques beyond the scope of this work). Examples are \emph{time period}, \emph{subjective sensitivity}, and \emph{visits}.

\textbf{\textit{Time period.}} The time period that a trajectory $\traj$ was taken,e.g., weekdays or weekend, can be a factor contributing to its VOI. For example, a person may often have commute-related trajectories during weekdays but more leisure-related trajectories on weekends. While time period does not represent actual locations, it can be used in conjunction with other methods, e.g., an owner can have different values for trajectories during weekdays compared to weekends because of privacy concerns. 

\textbf{\textit{Subjective sensitivity.}}
Each person may have their own sensitivity for different types of locations, which may lead to different sensitivity for different trajectories. For example, one may feel their workplace is more sensitive than their favorite coffee shop, thus having a higher sensitivity for the trajectory from home to work than the one to the coffee shop. While this information can contribute to the VOI of a trajectory, incorporating it requires additional information about the location measurements, which is not the focus of this work. When this information is available, it can be used in conjunction with the proposed methods in this work to better quantify the VOI tailored to the owner's subjective reasoning.

\textbf{\textit{Visits.}}
This method bases on the number of visits of a trajectory to quantify the VOI. Its main drawback is how to define a visit. It is often not feasible for the owner to manually define all visits for all of their trajectories. On the other hand, complicated techniques to automatically find visits~\cite{yue2019detect} are often used to segment a long sequence of measurements into trajectories, which is beyond the scope of this work. These techniques also often require additional information and/or a complex set of parameters where a small change of some parameters may result in a significantly different number of visits, which is not desirable.

\section{The Information Gain Framework}
\label{sec:ig_based}
This section describes the proposed framework to quantify the VOI of a trajectory. The framework is based on the notion of \emph{information gain (IG)} which quantifies the reduction of uncertainty when new information is available. Adopting IG to trajectories is not straightforward, and we enable it by  transforming each trajectory to a canonical representation. %
This transformation is done by employing a reconstruction method that can produce continuous-time probabilistic predictions along the trajectory. The Gaussian process (GP) is used in this work as the reconstruction method and discussed in the next section, but we emphasize that our IG framework accepts any reasonable method of producing probabilistic location inferences.

\subsection{Trajectory Information Gain}
We define information gain $\ig_{\igperiod}(\trajdegraded, \priorinfo)$ of a degraded version $\trajdegraded$ of a trajectory $\traj$ as the total reduction of uncertainty about locations of the owner over time period $\igperiod$ compared to the uncertainty from the prior knowledge $\priorinfo$. For a typical human trajectory, there are potentially several reasonable choices for $\igperiod$. In this work, $\igperiod$ is defined as the entire day covering the trajectory, because a typical trajectory would not extend beyond a day. We propose $\ig_{\igperiod}(\trajdegraded, \priorinfo)$ to quantify the intrinsic VOI of $\traj$ and/or its degraded version $\trajdegraded$. This section describes how $\ig_{\igperiod}(\trajdegraded, \priorinfo)$ is derived. In Section~\ref{sec:eval}, we will explain how $\ig_{\igperiod}(\trajdegraded, \priorinfo)$ satisfies almost all of the criteria in Table~\ref{tbl:methods_characteristics}, making it a better choice than the previous methods we described.

Recall that the owner can assume that the recipient obtained some prior knowledge $\priorinfo$, e.g., from public data or from previous noisier release of the same trajectory $\traj$. From $\priorinfo$, the owner can assume that the recipient can derive a prior probability distribution $\probdist(\dpoint | \priorinfo)$ for the location of the owner at a specific time $\dpoint.\dpointtime$. 

The owner can then assume that after receiving $\trajdegraded$, the recipient can use a model to reconstruct (or predict/interpolate) locations of the owner in continuous time. This assumption is made, because without knowledge of the recipient's intent, the owner should act conservatively and assume the recipient will exploit the new data fully, such as with a maximally accurate reconstruction. Such an inference can be represented as a posterior distribution $\probdist(\dpoint | \trajdegraded, \priorinfo)$.

Subsequently, the information gain $\ig_{\dpointtime}(\trajdegraded, \priorinfo)$ at timestamp $\dpointtime = \dpoint.\dpointtime$ indicates the reduction of uncertainty from $\probdist(\dpoint| \priorinfo)$ to $\probdist(\dpoint | \trajdegraded, \priorinfo)$. The uncertainty is measured by \emph{differential entropy}~\cite{cover2012differentialentropy} (or continuous entropy), because both the prior and posterior distributions are likely continuous distributions in space. 

The differential entropy $\diffentropy(X)$ of a random variable $X$ with probability density function $\probdist$ whose support is a set $\dset$ is defined as
\begin{equation}
    \diffentropy(X) = - \int_{\dset} \probdist(x) \log \probdist(x) dx
\label{eq:diffentropy}
\end{equation}
Several popular probability distributions have a closed-form expression for their differential entropy, e.g, if a $X$ is a Gaussian random variable with distribution $\mathcal{N}(\mu, \sigma^2)$, its differential entropy is
\begin{equation}
    \diffentropy(X) = \frac{1}{2} \log 2 \pi e \sigma^2
\end{equation}
The entropy values of latitude and longitude of $\dpoint$ are calculated separately using Equation~\ref{eq:diffentropy}, and summed to get $\diffentropy(\dpoint | \priorinfo)$ and $\diffentropy(\dpoint | \trajdegraded, \priorinfo)$.

Thus the IG at time $t$,  $\ig_{\dpointtime}(\trajdegraded, \priorinfo)$, can be calculated as
\begin{equation}
    \ig_{\dpointtime}(\trajdegraded, \priorinfo) = \diffentropy(\dpoint | \priorinfo) - \diffentropy(\dpoint | \trajdegraded, \priorinfo)
\end{equation} 

\begin{figure}[htbp!]
\centering
\begin{subfigure}{.5\linewidth}
  \centering
  \includegraphics[width=\linewidth]{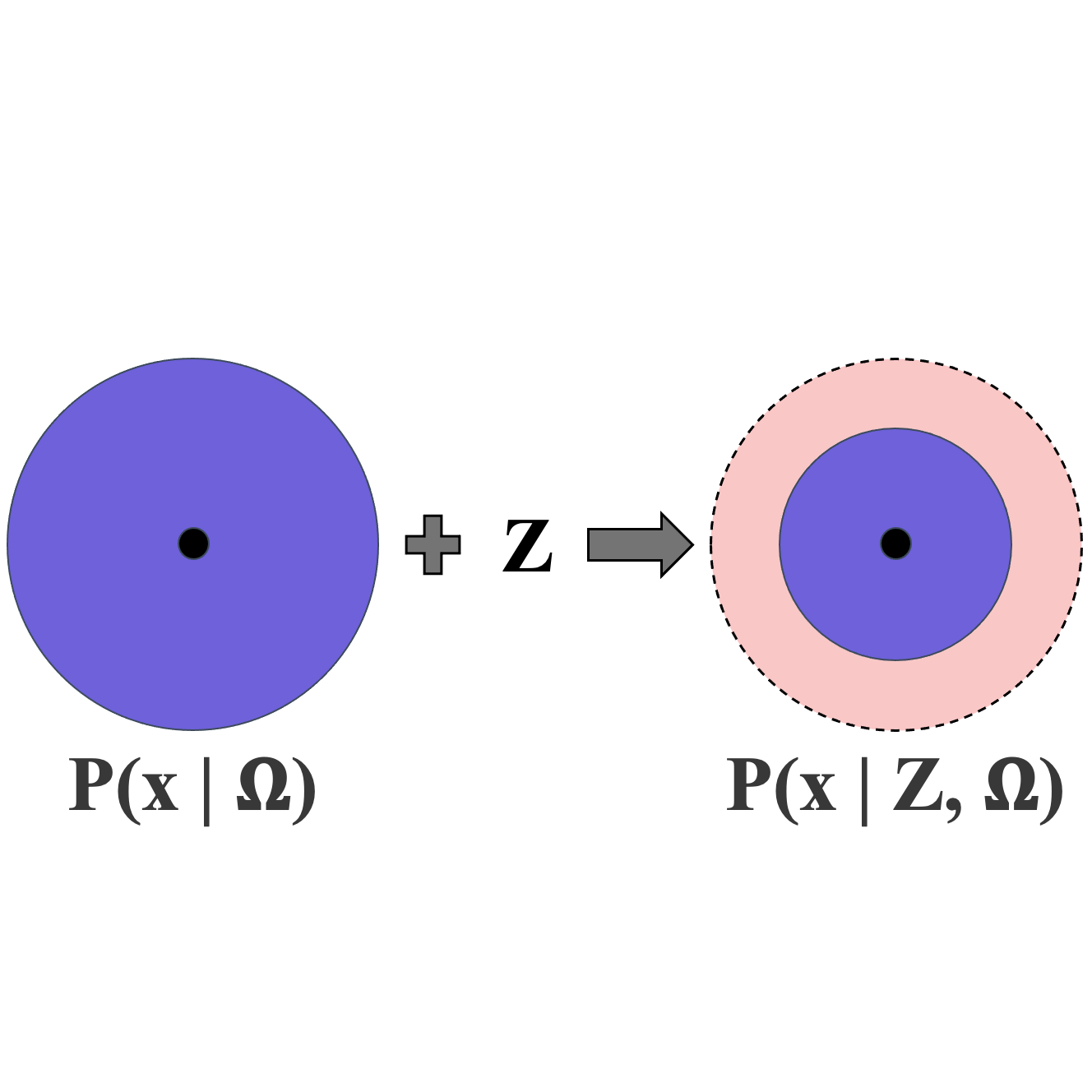}
  \caption{$\ig_{\dpointtime}(\trajdegraded, \priorinfo)$ illustration}
  \label{fig:ig_illustration_point}
\end{subfigure}%
\begin{subfigure}{.5\linewidth}
  \centering
  \includegraphics[width=\linewidth]{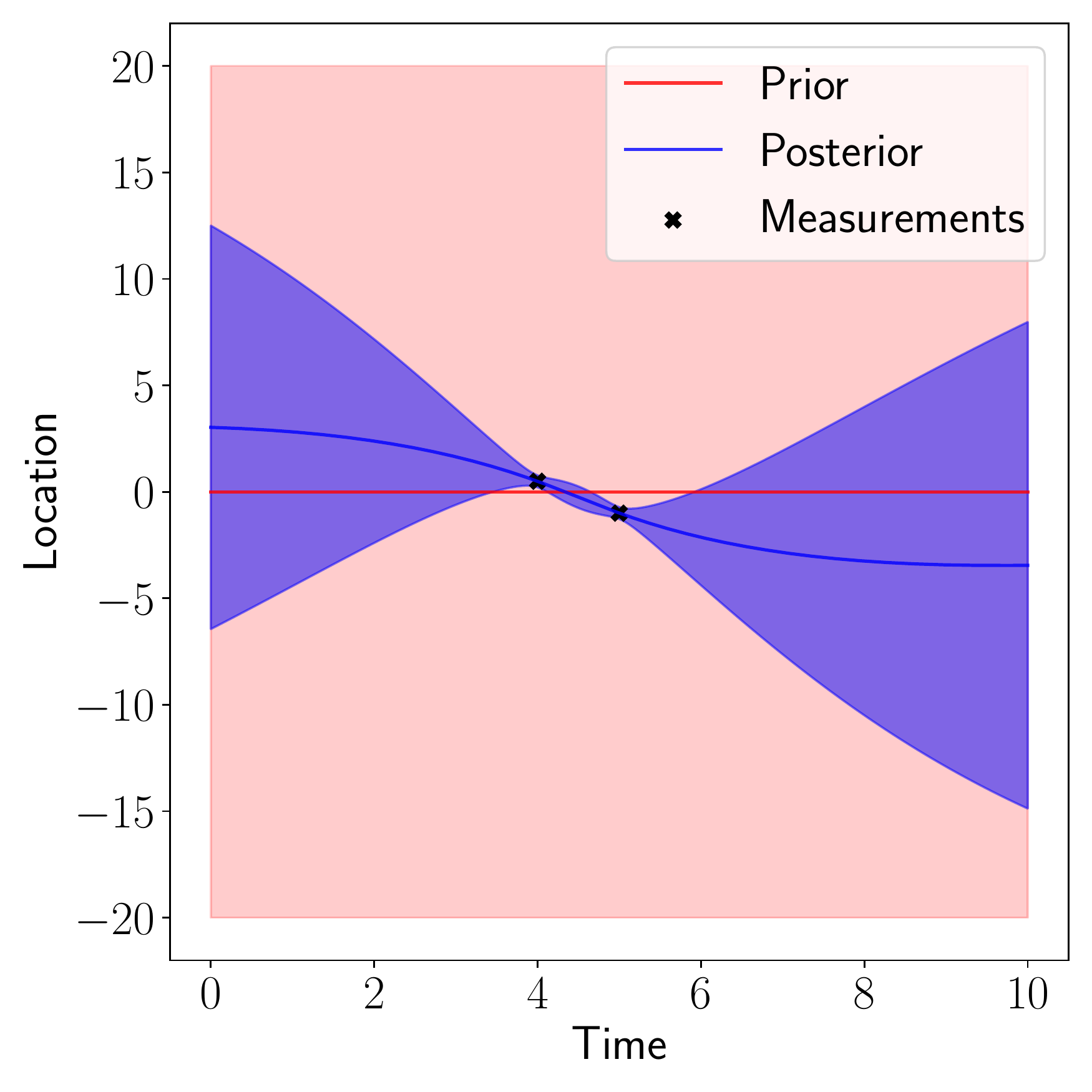}
  \caption{$\ig_{\igperiod}(\trajdegraded, \priorinfo)$ illustration}
  \label{fig:ig_illustration_trajectory}
\end{subfigure}
\caption{Illustrations of information gain (\protect\subref{fig:ig_illustration_point}) at a single timestamp and (\protect\subref{fig:ig_illustration_trajectory}) over a time period. The reduced uncertainty is shown in the red area minus the blue area.} 
\label{fig:ig_illustration}
\end{figure}

Figure~\ref{fig:ig_illustration_point} illustrates $\ig_{\dpointtime}(\trajdegraded, \priorinfo)$ where the black dot shows the prediction mean and the blue area shows the uncertainty as the circle with a radius which is twice the standard deviation of the predicted distribution $\probdist(\dpoint | \priorinfo)$. If $\trajdegraded$ is available, the uncertainty is reduced to the smaller blue circle on the right representing $\probdist(\dpoint | \trajdegraded, \priorinfo)$. $\ig_{\dpointtime}(\trajdegraded, \priorinfo)$ quantifies the reduction and is illustrated as the red ring.

Note that $\dpoint$ does not need to be in any trajectory, because the probabilistic reconstruction is continuous in time. For example, $\dpoint$ can be in between two measurements $\dpoint_i, \dpoint_{i+1} \in \traj$, thus, knowing $\dpoint_i$ and $\dpoint_{i+1}$ would help reduce the uncertainty of where $\dpoint$ can be. 

Subsequently, the information gain $\ig_{\igperiod}(\trajdegraded, \priorinfo)$ over a time period $\igperiod$ can be calculated by integrating $\ig_{\dpointtime}(\trajdegraded, \priorinfo)$ for all time $\dpointtime \in \igperiod$, i.e., 
\begin{equation}
    \ig_{\igperiod}(\trajdegraded, \priorinfo) = \int_{\dpointtime \in \igperiod} \ig_{\dpointtime}(\trajdegraded, \priorinfo) d \dpointtime
\end{equation}

Figure~\ref{fig:ig_illustration_trajectory} illustrates $\ig_{\igperiod}(\trajdegraded, \priorinfo)$ computed over a time period $\igperiod = [0, 10]$ with prior $\probdist(\dpoint | \priorinfo) = \mathcal{N}(0, 100)$ for each timestamp. Over this period, the red line shows the prediction mean of the prior distributions, and the red area shows the prior uncertainty. When $\trajdegraded$, shown as two black crosses, is available, the uncertainty is reduced to the blue area. The amount of such reduction is shown as the red area minus the blue area and is quantified using $\ig_{\igperiod}(\trajdegraded, \priorinfo)$.

\subsection{Reconstruction Method}
\label{sec:reconstruction_method}
To compute $\ig_{\igperiod}(\trajdegraded, \priorinfo)$, a probabilistic reconstruction method is needed to reconstruct locations of the owner given $\priorinfo$ and/or $\trajdegraded$. 
A Gaussian process (GP) is used in this work as the reconstruction method because of its flexibility to incorporate different types of information. However, we emphasize that reconstruction is not the focus of this paper. The IG framework accepts any reasonable probabilistic reconstruction method. Thus, we briefly discuss important aspects of the GP. More details about GPs can be seen in~\cite{Rasmussen2005GPM}. 

For a scalar function $f(t)$, a GP implies that any subset of points sampled from $f(t)$ is distributed according to a multidimensional Gaussian. Two independent GPs are created for longitude and latitude prediction. The input for a GP are pairs $<t, f(t)>$ where $f(t)$ is longitude or latitude of the measurement at time $t$. The output for a set of timestamps $\gppredset^*$ are predictions $f^*(t)$ for each $t \in \gppredset^*$. 

A GP depends on a scalar covariance kernel/function $k(t, t')$ defining how much a measurement at time $t$ correlates with a measurement at time $t'$. In general, the correlation decreases to zero as $|t-t'|$ gets larger. A good kernel can help incorporate different types of information into the model, especially when kernels can be combined together. In our implementation, two common kernels are summed together: the main kernel is a Matérn kernel that captures the relationship between measurements as well as the prior knowledge, and a white kernel to capture measurement uncertainty.

The formula of the Matérn kernel used in this work is
\begin{equation}
    k_{\textit{Matérn}}(t, t') = \dpointsigma_f^2\Big(1 + \frac{\sqrt{3}}{l}|t - t'|\Big)\exp\Big(-\frac{\sqrt{3}}{l}|t - t'|\Big)
\end{equation}
Setting $\dpointsigma_f$ to the standard deviation of the prior distribution can help capture the prior knowledge, so that when the model becomes more uncertain, the standard deviation of the prediction will gradually reach this value. The length scale $l$ is trained from data.

The formula of a white kernel is
\begin{equation}
    k_{white}(t, t') = \begin{cases}
       \dpointsigma_m^2 &\quad\text{if t = t'}\\
       0 &\quad\text{otherwise}
     \end{cases}
\end{equation}
Setting $\dpointsigma_m$ to $\dpoint.\dpointsigma$ or $\dpointdegraded.\dpointsigma$ helps capture the measurement uncertainty. 

The final kernel is
\begin{equation}
    k(t, t') = k_{\textit{Matérn}}(t, t') + k_{white}(t, t')
\end{equation}

Finally, a GP also depends on a mean function $m(t)$ which defines the expected mean values of the measurements. In our adaption, the mean function $m(t)$ of a GP is set to be the regression line obtained by running a linear regression on the data in $\priorinfo$.

\section{Evaluations of the IG framework}
\label{sec:eval}
This section provides an evaluation of how the IG framework can capture each characteristic from Table~\ref{tbl:methods_characteristics}, both qualitatively and quantitatively. The quantitative evaluation is performed on the Geolife dataset, which is a large, real-world trajectory dataset. 

\subsection{Dataset}
\label{sec:datasets}
The Geolife dataset~\cite{zheng2009mining} is used for experiments to quantitatively justify claims in the paper. This is a trajectory dataset collected by 182 people carrying GPS loggers and GPS-phones in the Beijing area from April 2007 to August 2012. A trajectory is represented by a sequence of measurements containing the latitude, longitude, and timestamp, with a variety of sampling rates. With this large, real-world dataset, covering a large span in both space and time, different types of devices, and a variety of sampling rates, we expect it is representative of the observations in other real-world datasets. 

Measurements in the Geolife dataset are filtered to retain only ones within the Bejing area with longitude from $116.20$ to $116.55$ degrees and latitude from $39.80$ to $40.06$ degrees. This area covers almost all measurements and, in total, more than 16 million measurements were retained from all 182 people. The latitude/longitude coordinates in each measurement are converted to local Euclidean coordinates $(x, y)$ in meters with the reference origin $(lon_0, lat_0)$ arbitrarily chosen as the center of the aforementioned area. In local Euclidean coordinates, the area has a lower left coordinate of $(-15000, -15000)$ and an upper right coordinate of $(15000, 15000)$. 

Measurements of each individual are separated into trajectories by iterating through the ordered measurements and creating a new trajectory whenever the time gap between the current measurement and the next measurement is more than $\trajtimegap$ seconds. 
The maximum time gap $\trajtimegap$ is arbitrarily chosen at $300$ seconds or $5$ minutes. The actual value of $\trajtimegap$ does not significantly change the experimental results. We also emphasize that finding trajectories is not the focus of this work. %
Thus, more sophisticated methods to find trajectories from measurements can also be used as an alternative to this segmentation approach. The final dataset has 45,831 trajectories.

\subsection{Experiment Setup}
\label{sec:exp_setup}

\label{sec:ig_and_char_from_raw_trajectory}
\begin{figure*}[htbp!]
\centering
\begin{subfigure}{0.166\linewidth}
  \centering\captionsetup{justification=centering}
  \includegraphics[width=\textwidth]{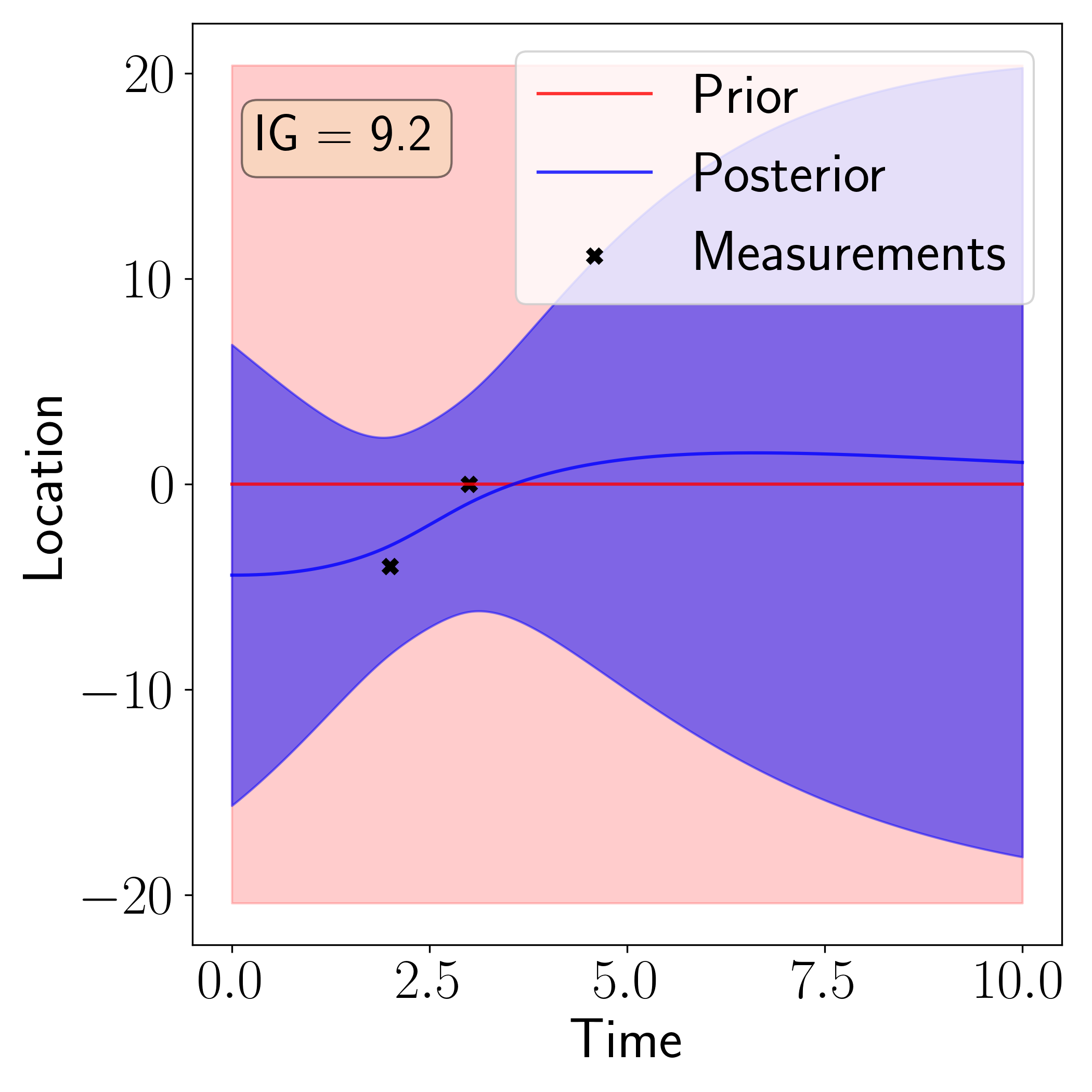}
  \caption{Small size,\\ large uncertainty}
  \label{fig:IG_qualitative_2_points_noisy}
\end{subfigure}%
\begin{subfigure}{.166\linewidth}
  \centering\captionsetup{justification=centering}
  \includegraphics[width=\textwidth]{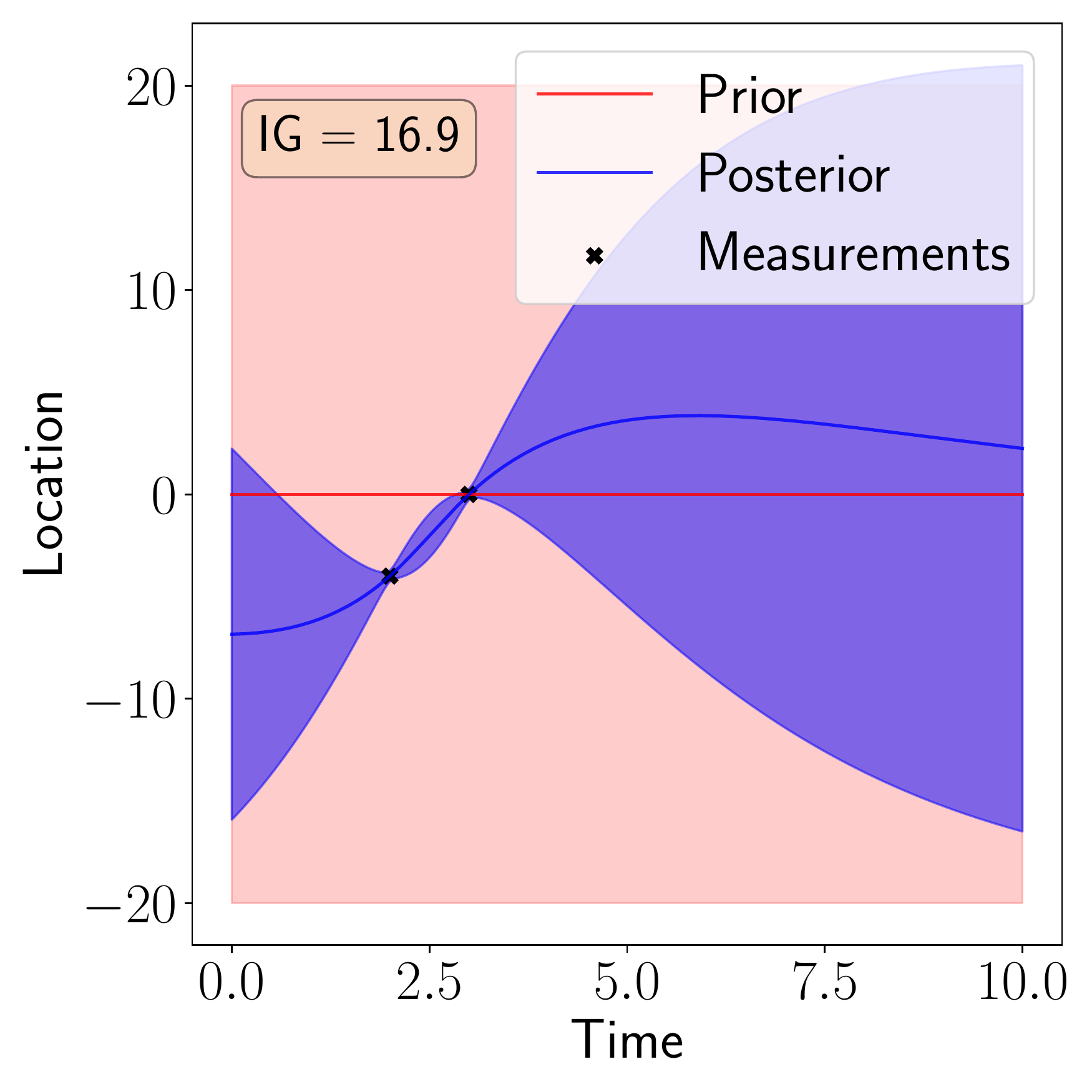}
  \caption{Small size,\\ small uncertainty}
  \label{fig:IG_qualitative_2_points}
\end{subfigure}%
\begin{subfigure}{.166\linewidth}
  \centering\captionsetup{justification=centering}
  \includegraphics[width=\textwidth]{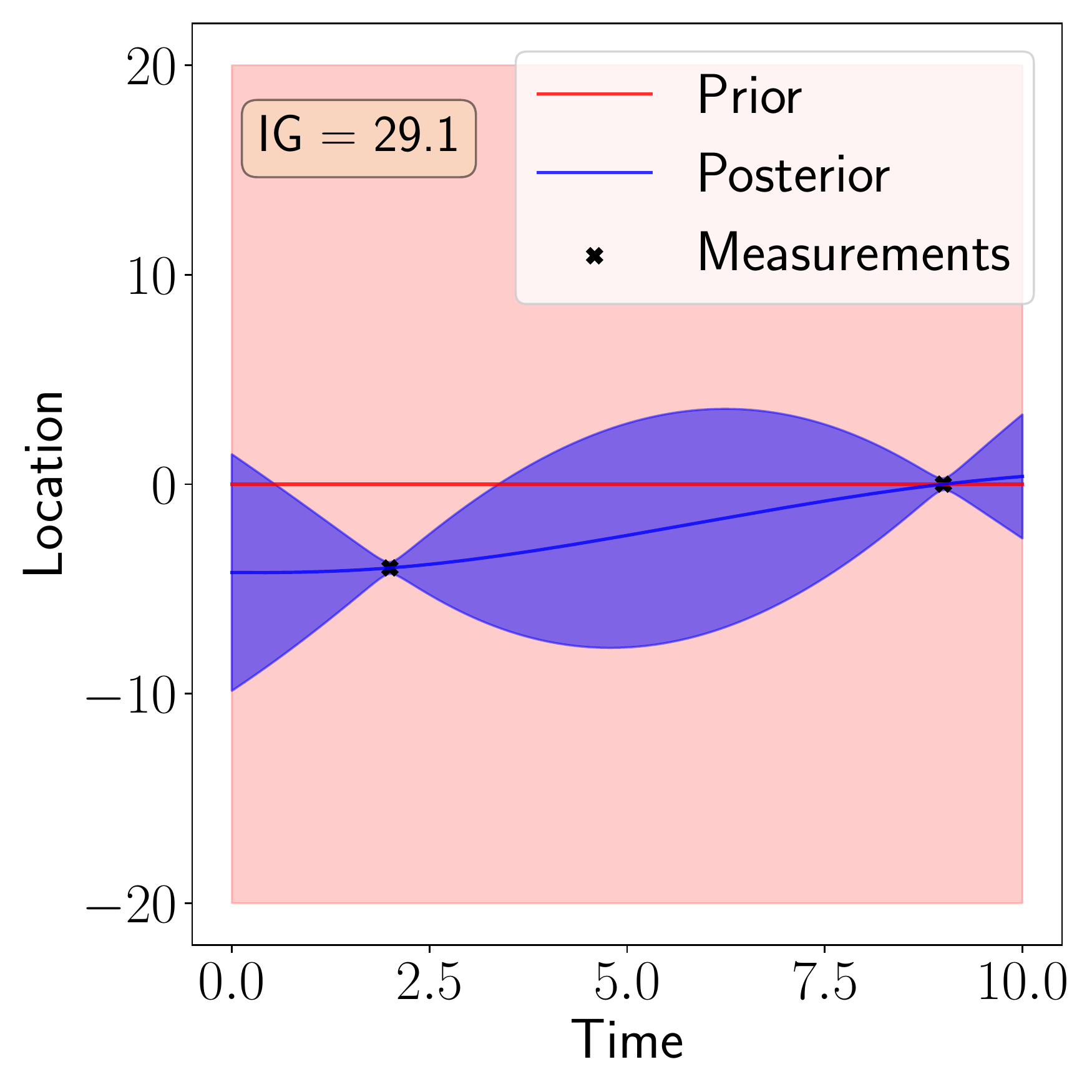}
  \caption{Small size,\\ long duration}
  \label{fig:IG_qualitative_2_points_far}
\end{subfigure}%
\begin{subfigure}{.166\linewidth}
  \centering\captionsetup{justification=centering}
  \includegraphics[width=\textwidth]{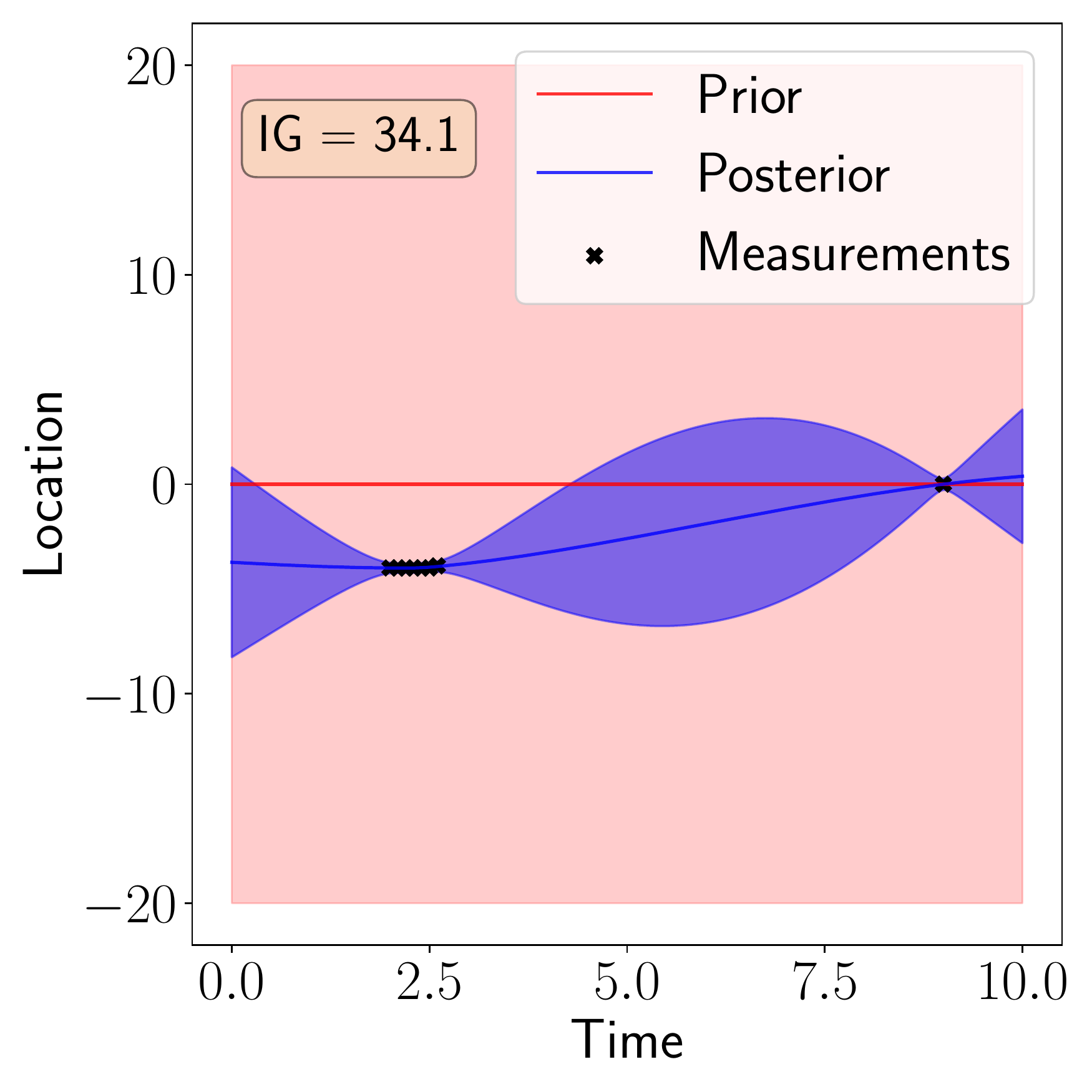}
  \caption{Large size, clustered\\ around 1st point}
  \label{fig:IG_qualitative_8_points_close}
\end{subfigure}%
\begin{subfigure}{.166\linewidth}
  \centering\captionsetup{justification=centering}
  \includegraphics[width=\textwidth]{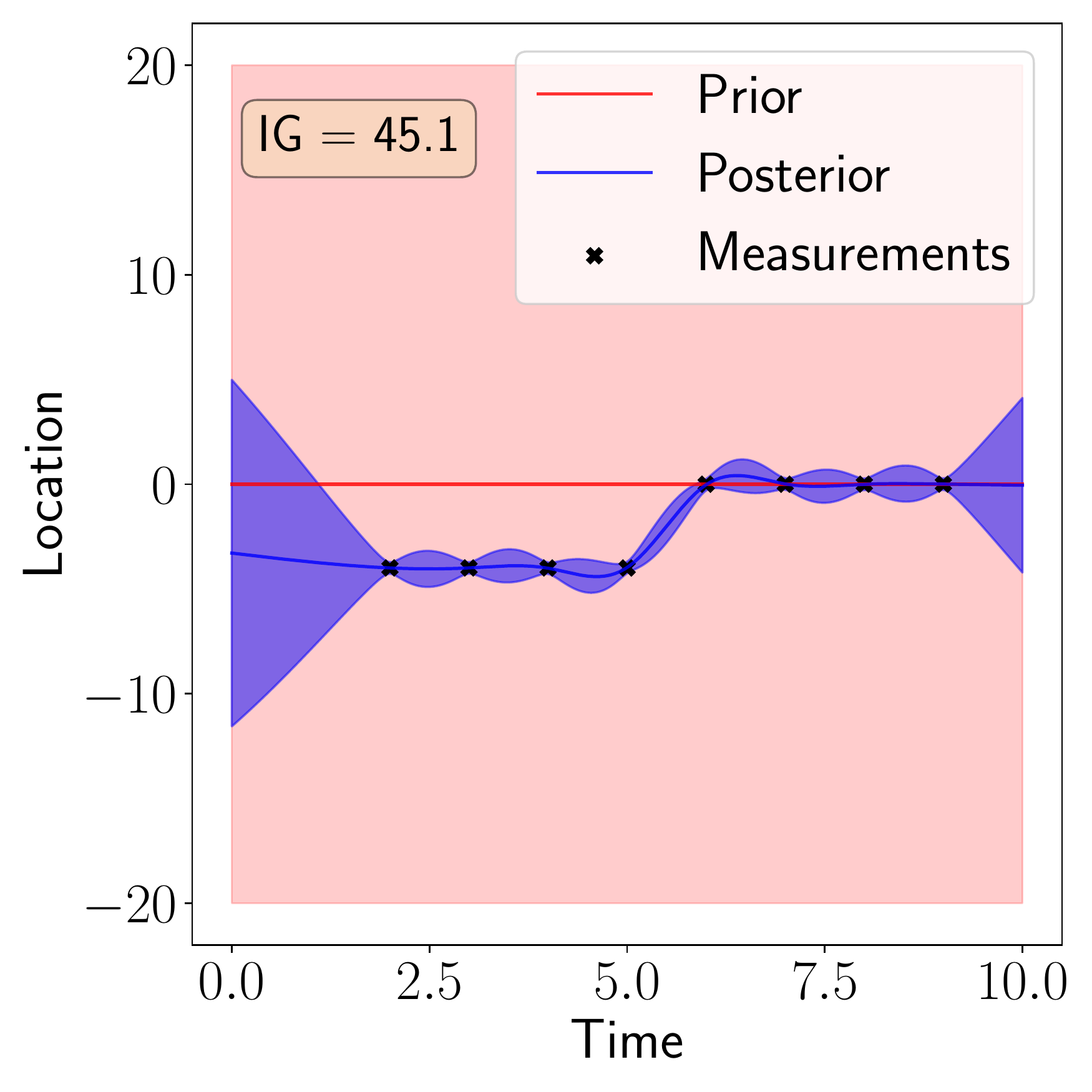}
  \caption{Same size as (\protect\subref{fig:IG_qualitative_8_points_close}),\\ scattered}
  \label{fig:IG_qualitative_8_points_dispersed}
\end{subfigure}%
\begin{subfigure}{.166\linewidth}
  \centering\captionsetup{justification=centering}
  \includegraphics[width=\textwidth]{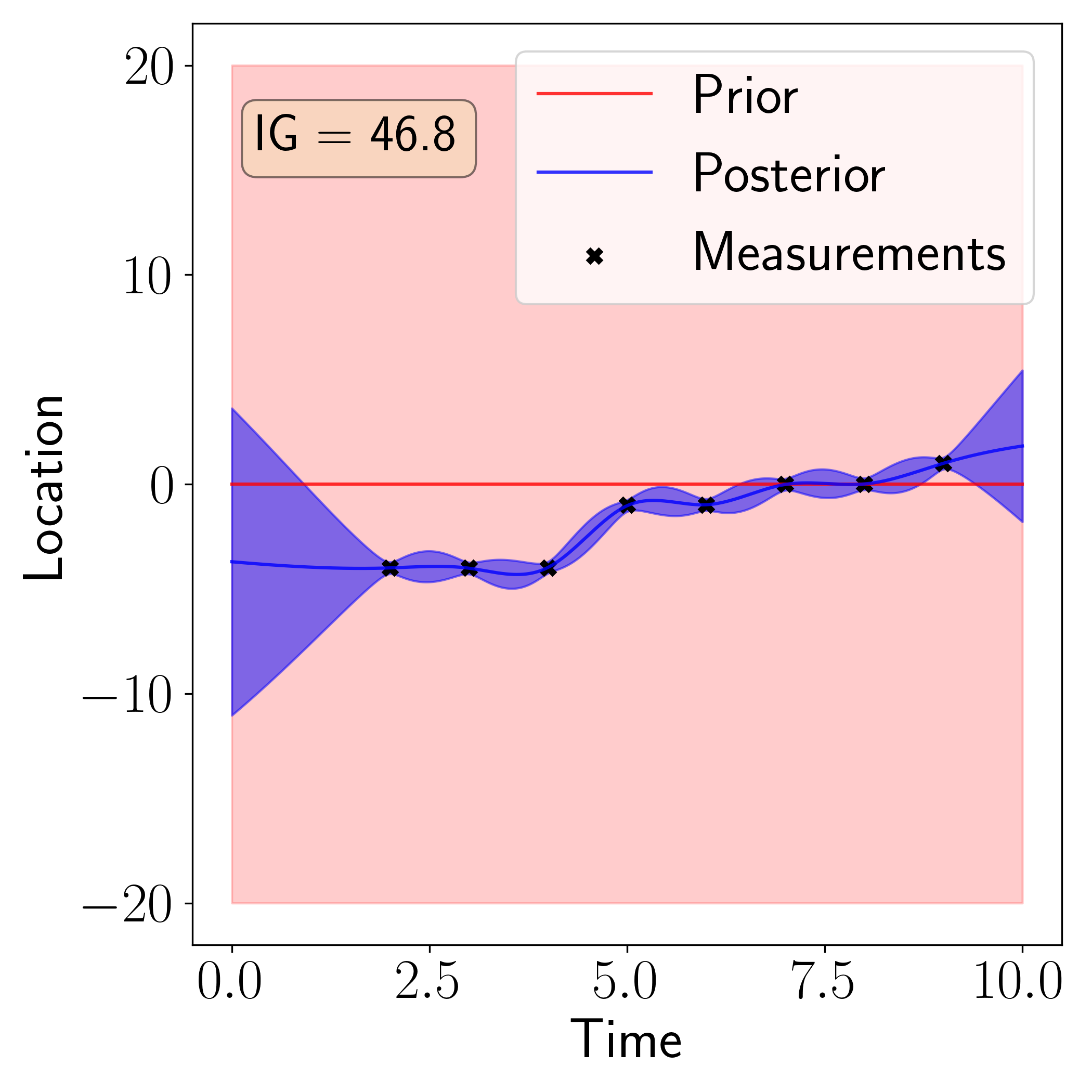}
  \caption{Same size and times as (\protect\subref{fig:IG_qualitative_8_points_dispersed}), more locations}
  \label{fig:IG_qualitative_8_points_dispersed_spatial}
\end{subfigure}
\caption{An illustration of how IG can capture trajectory characteristics. The figures from left to right showing the effect of measurement uncertainty, duration, size, temporal, and spatial distribution of a trajectory.}
\label{fig:IG_qualitative}
\end{figure*}

The experiments were conducted on the aforementioned dataset with various sets of parameters. The results, such as the correlation coefficient or regression line, are computed from all trajectories. Several outliers are removed for visual presentation purpose but still included in all computations. The default size of grid cells/temporal bins to calculate spatial/temporal entropy is $10$~meters/one minute.%

The raw measurements do not include uncertainty. However, because they were recorded by GPS-equipped devices and the noise for GPS-equipped smartphones is about $3$~meters~\cite{gpsaccuracy}, each measurement is assumed to have $\dpoint_i.\dpointsigma = 3$.
For perturbation degradation, the noise is added such that the total noise (i.e., $\dpointdegraded_i.\dpointsigma$) is in $\{3, 10, 100, 200, 300, 400\}$~meters. 
The ratios for truncation and subsampling degradation (i.e., $\truncationratio$ and $\subratio$) are $\{0.8, 0.6, 0.4, 0.2, 0.05\}$, which mean $\{80\%, 60\%, 40\%, 20\%, 5\%\}$ of the raw trajectory. The subsampling is implemented so that the subsampled data with a higher ratio is a superset of the subsampled data with a smaller ratio.

A measurement is assumed to be inside the Beijing area. This choice is reasonable for this dataset. However, it is not crucial to the framework where any reasonable prior knowledge can be used (e.g., measurements can be anywhere on Earth) and would only change the scale of IG. This prior knowledge is represented by a Gaussian distribution with extremely high variance where the mean is at the center of the area (i.e., coordinates $(0, 0)$) and the standard deviation is $\dpointsigma_0 = 7500$ meters (i.e., the distance from the center to an edge of the area is twice this standard deviation). This is an \emph{uninformative} prior and called the \emph{Gaussian prior}, i.e., for each timestamp, the location is assumed to be $\mathcal{N}(0, \dpointsigma_0^2)$ for each dimension.

It is also natural to consider any previously released data as prior knowledge when computing the VOI of the data the owner still retains. So we also consider previous releases as priors. This also demonstrates an advantage of the IG framework, where any previous release can be naturally considered when computing the VOI of the owner's remaining data. To illustrate these cases, depending on the nature of a degradation, different \emph{informative} priors based on previous releases are considered. 
For perturbation, two additional priors are \emph{$400$m noise} and \emph{$300$m noise} priors, illustrating the cases where the owner released their trajectories at those noise levels before. For example, with the $400$m noise prior, when computing the IG for a degraded version $\trajdegraded_{100}$ with $100$m noise of a trajectory $\traj$, the owner may consider the prior knowledge is the degraded version $\trajdegraded_{400}$ with $400$m noise of $\traj$. Thus, there are three priors for perturbation in total: the uninformative \emph{Gaussian prior} and two informative priors \emph{$400$m noise} and \emph{$300$m noise}. 
Similarly, there are three priors for truncation/subsampling: the uninformative \emph{Gaussian prior} and two informative \emph{$5\%$ trajectory} and \emph{$20\%$ trajectory} priors, illustrating the cases when the owner released trajectories with $0.05$ or $0.2$ truncation/subsampling ratio before. 
Other prior knowledge, e.g., previously purchased trajectories, can also be considered as priors, provided that they can be expressed as prior location distributions.

The GP is trained on the prior $\priorinfo$ (e.g., 300m prior), except when the prior is the uninformative Gaussian prior, when it is then trained on the new degraded version $\trajdegraded$. The degraded version $\trajdegraded$ and prior $\priorinfo$ are combined and then provided to the GP. For perturbation, measurements $\trajdegraded$ and $\priorinfo$ of each timestamp are combined using inverse-variance weighting~\cite{hartung2011statistical}. For truncation and subsampling, because the higher-ratio degraded version is always a superset of a lower-ratio degraded version (e.g, $\trajdegraded_{5\%} \subset \trajdegraded_{20\%} \subset \trajdegraded_{40\%}$), the combined data is $\trajdegraded$. This is a reasonable approach for GP. There can be other methods to better combine a prior and new data, especially if another prediction model is used instead of GP; however, this is not the focus of the paper. While we expect similar trends, especially from models similar to GP such as a Kalman filter~\cite{kalman1960new} or a particle filer~\cite{doucet2001introduction}, absolute values from other models can be slightly different.

For GP kernels, $\dpointsigma_0 = 7500$ is provided to the Matérn kernel. Total noise $\dpointdegraded.\dpointsigma$ is provided to the white kernel. Length scale $l$ can take values in $[0.01, 10]$ and is trained separately for each degraded version. IG over the time period is computed using numerical integration with the trapezoid rule. Finally, logarithms are base 2.

\subsection{IG Capturing Trajectory Characteristics}

For qualitative evaluation, Figure~\ref{fig:IG_qualitative} illustrates the IG for different cases in one spatial dimension. Starting from left to right with a trajectory having two noisy measurements close to each other in Figure~\ref{fig:IG_qualitative_2_points_noisy}, the IG increases as expected when the measurements are less noisy as in Figure~\ref{fig:IG_qualitative_2_points}. When the duration increases, illustrated by two measurements being farther apart in Figure~\ref{fig:IG_qualitative_2_points_far}, the IG also increases because these two measurements can also help \emph{reduce uncertainty for the longer time in between these two measurements}. When more measurements are available as in Figure~\ref{fig:IG_qualitative_8_points_close}, these points help reduce more uncertainty around them, thus increasing the IG further. The IG keeps increasing when the measurements are temporally distributed more evenly as shown in Figure~\ref{fig:IG_qualitative_8_points_dispersed} because, for most of the duration along the trajectory, there are temporally nearby measurements, thus helping reduce uncertainty. Finally, in Figure~\ref{fig:IG_qualitative_8_points_dispersed_spatial}, the IG also slightly increases when measurements appear in between the start and end locations (shown on the vertical axis), thus making the movement less abrupt. However, if the locations jump unpredictably, the uncertainty tends to increase compared to, e.g., when the person stays at the same place, thus decreasing the IG for that case. Therefore, the IG can only partially capture the spatial distribution and can capture the size, duration, temporal distribution, and measurement uncertainty, as reflected in Table~\ref{tbl:methods_characteristics}.

\begin{figure*}[htbp!]
\centering
\begin{subfigure}{.25\linewidth}
  \centering
  \includegraphics[width=\linewidth]{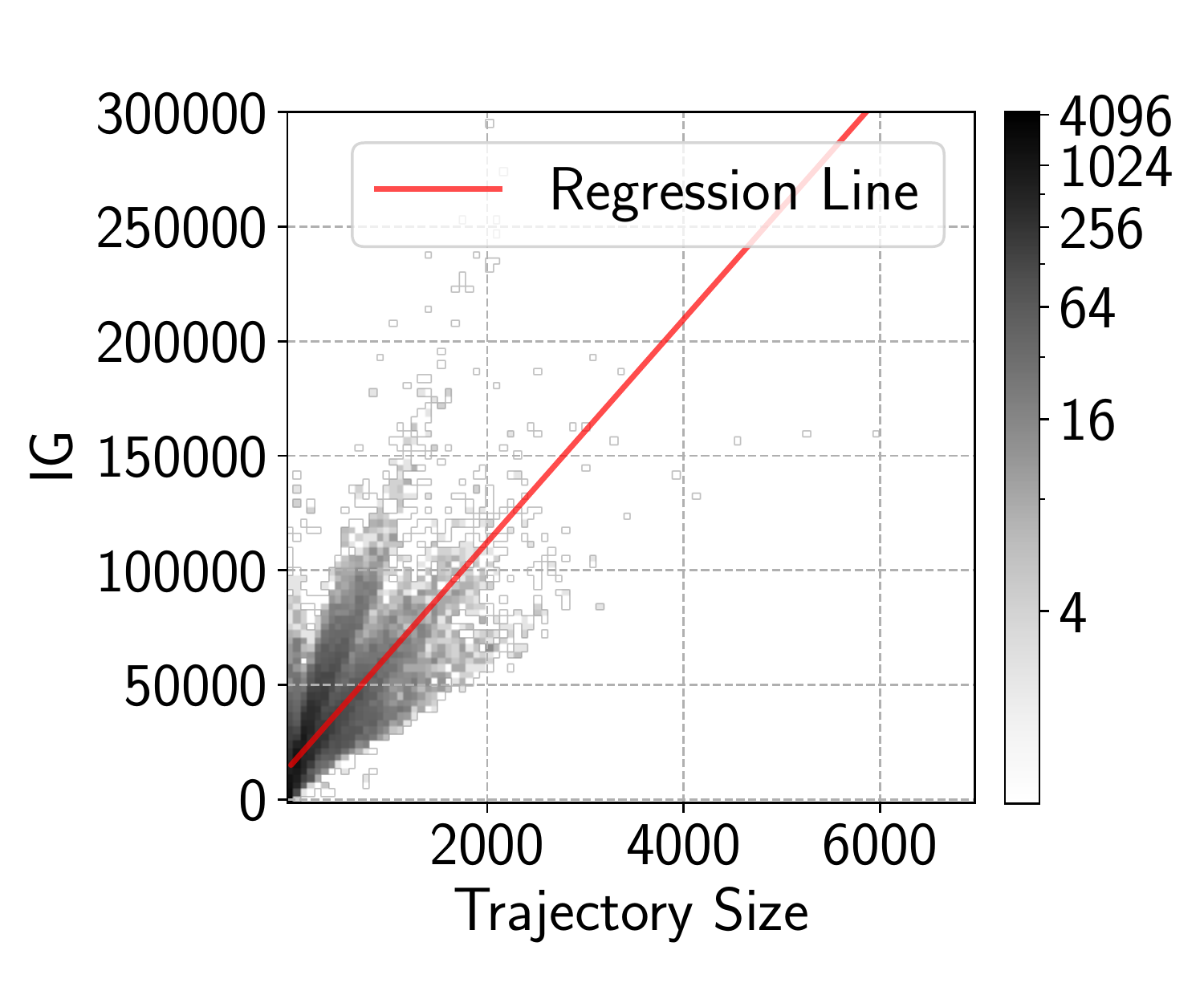}
  \caption{Size, $\corrcoeff = 0.85$}
  \label{fig:ig_vs_size_histogram}
\end{subfigure}%
\begin{subfigure}{.25\linewidth}
  \centering
  \includegraphics[width=\linewidth]{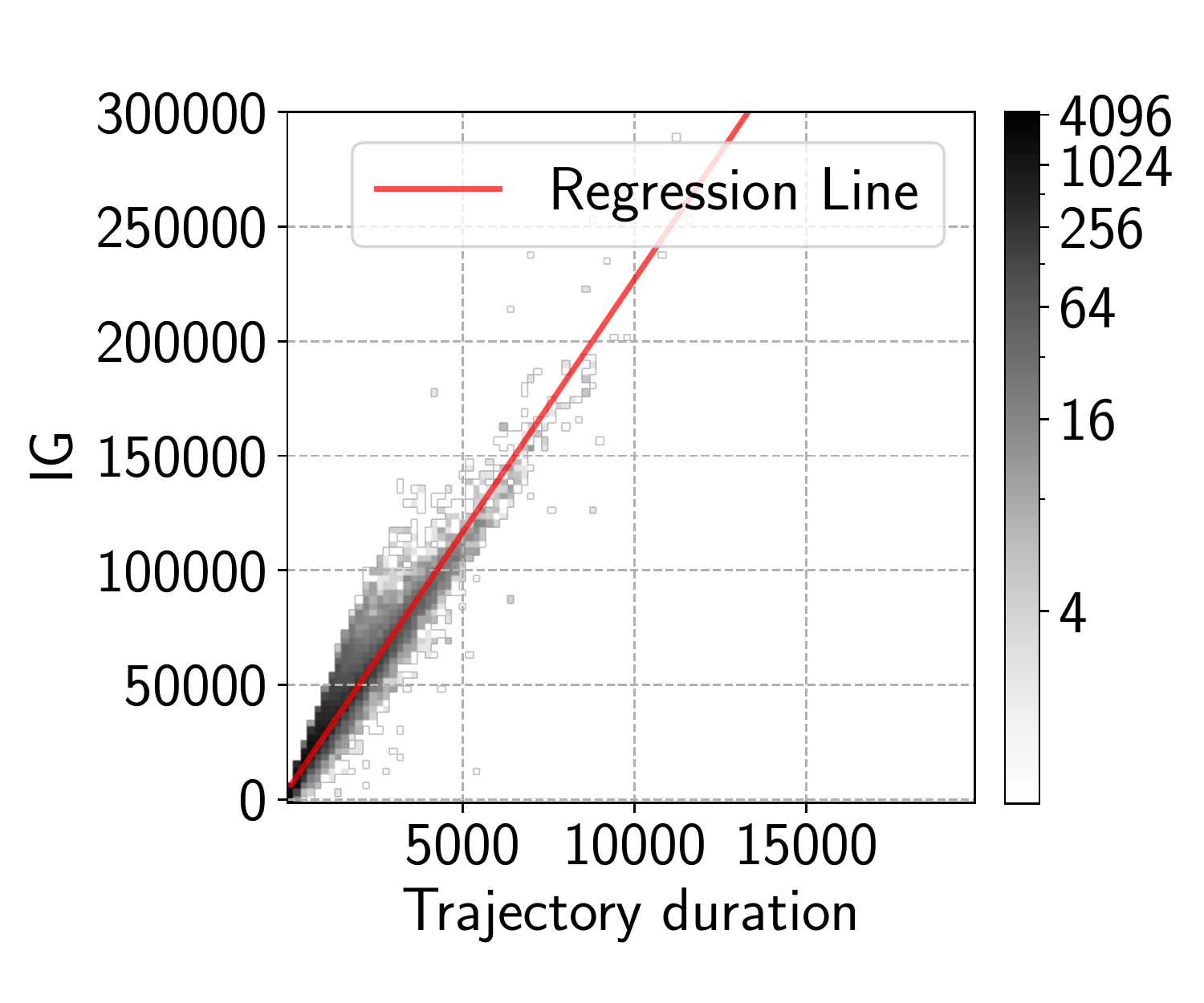}
  \caption{Duration, $\corrcoeff = 0.97$}
  \label{fig:ig_vs_duration_histogram}
\end{subfigure}%
\begin{subfigure}{.25\linewidth}
  \centering
  \includegraphics[width=\linewidth]{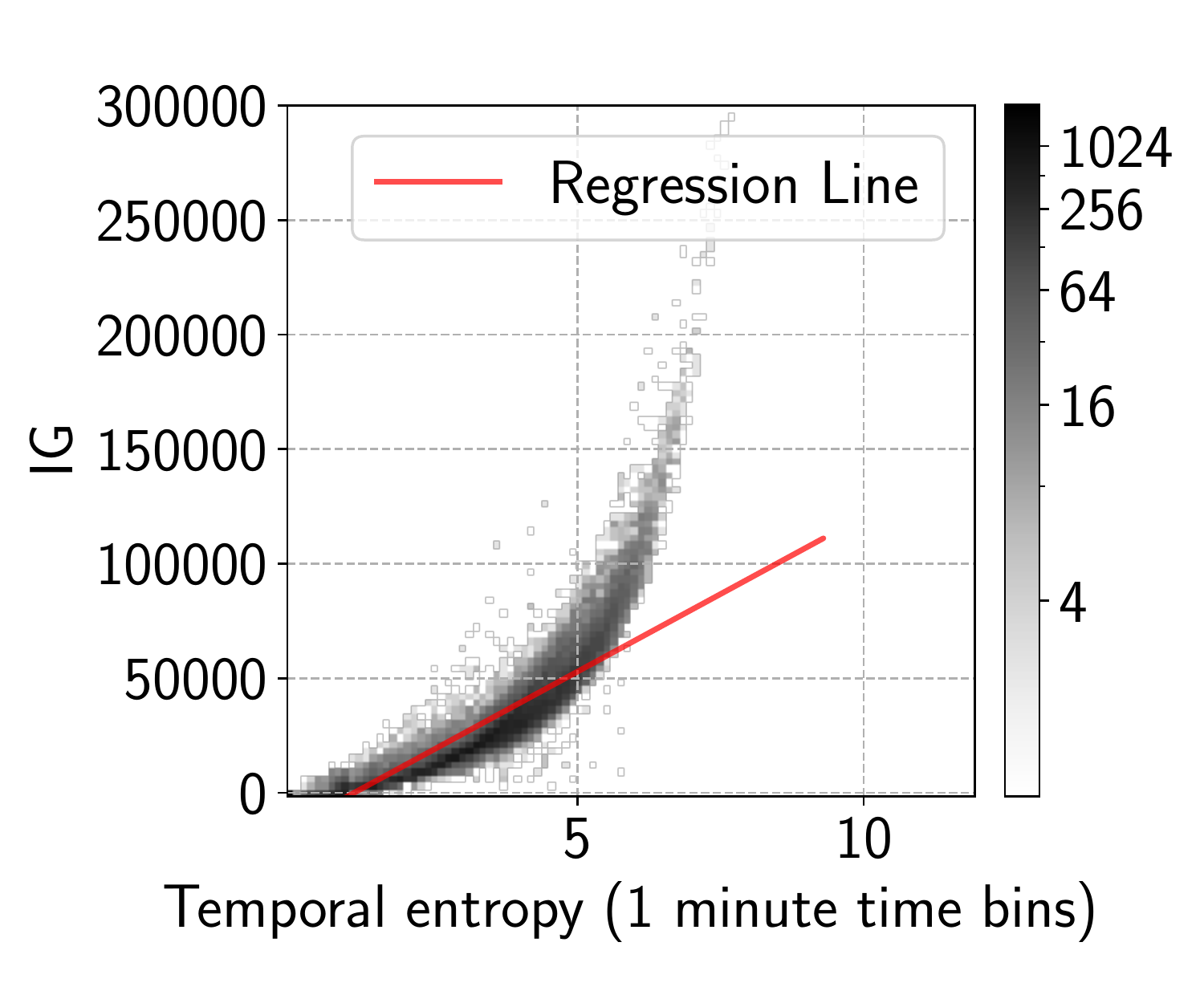}
  \caption{Temporal entropy, $\corrcoeff = 0.96$}
  \label{fig:ig_vs_temporal_entropy_histogram}
\end{subfigure}%
\begin{subfigure}{.25\linewidth}
  \centering
  \includegraphics[width=\linewidth]{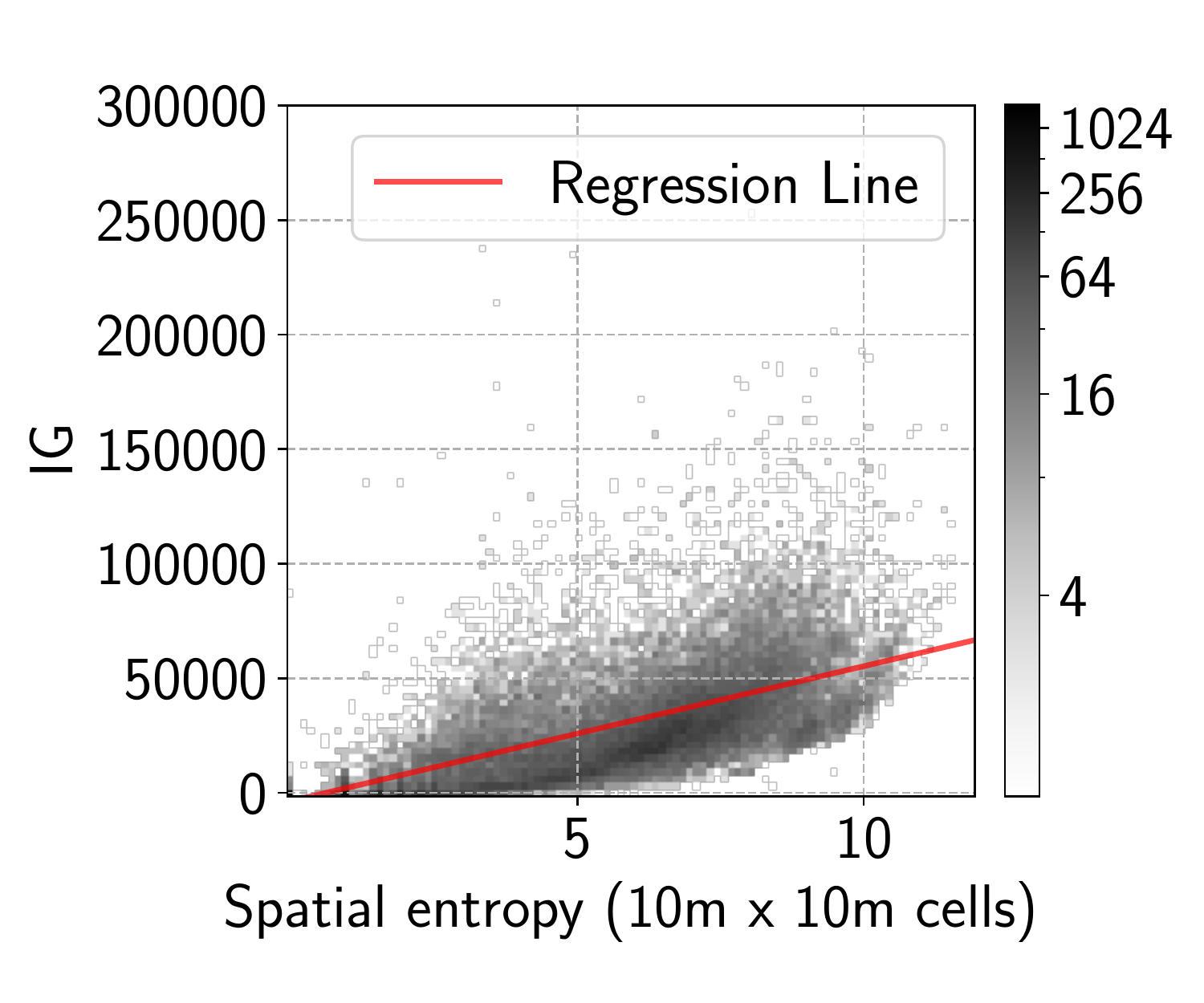}
  \caption{Spatial entropy, $\corrcoeff = 0.68$}
  \label{fig:ig_vs_spatial_entropy_histogram}
\end{subfigure}
\caption{Histogram of IG and (\protect\subref{fig:ig_vs_size_histogram}) size, (\protect\subref{fig:ig_vs_duration_histogram}) duration, (\protect\subref{fig:ig_vs_spatial_entropy_histogram}) spatial, (\protect\subref{fig:ig_vs_temporal_entropy_histogram}) temporal entropy, and their correlation coefficients.} 
\label{fig:ig_vs_char_raw_trajectory}
\end{figure*}

For quantitative evaluation using the Geolife data, the correlation coefficients of the IG with size, duration, and temporal entropy are $0.85, 0.97$ and $0.96$, respectively, which show very strong correlation. The correlation coefficients of the IG and spatial entropy is $0.68$ which is also close to a strong correlation. Figure~\ref{fig:ig_vs_char_raw_trajectory} shows the histograms of the IG and these characteristics on a log scale of the number of trajectories, along with a linear regression line indicating a clear trend between the IG and these characteristics. The regression uses Huber loss~\cite{huber1992robust} to mitigate the effect of outliers. Quantitative evaluation shown in Figure~\ref{fig:ig_noise} (explained in detail later) also shows that the IG can capture measurement uncertainty, because when a trajectory has large measurement uncertainty (i.e., less informative), the IG tends to decrease. Computing IG took around 2-4 seconds per trajectory on a single machine.

\subsection{IG Capturing Prior Knowledge and Degradation}
The impact of prior knowledge and degradation can be effectively captured by the IG framework. In general, when a trajectory is degraded more (e.g., larger total noise), the IG tends to decrease, matching the intuition that a lower quality trajectory gives lower information. Similar trends show in both the absolute value $\ig(\trajdegraded, \priorinfo)$ and the ratio $\frac{\ig(\trajdegraded, \priorinfo)}{\ig(\traj, \priorinfo)}$ of the IG compared to the raw trajectory $\traj$. 

\begin{figure}[htbp!]
\centering
\begin{subfigure}{.5\linewidth}
  \centering
  \includegraphics[width=\linewidth]{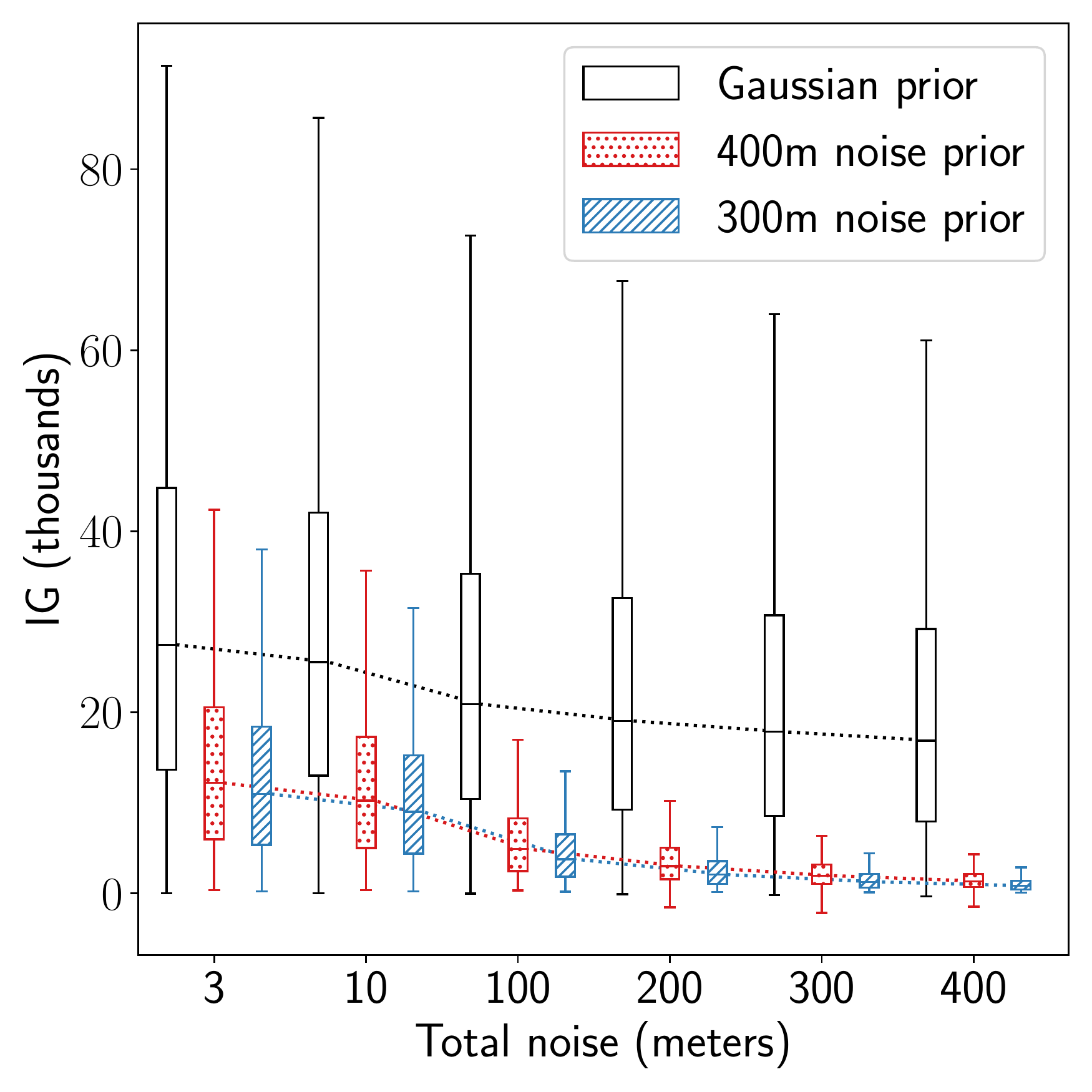}
  \caption{Absolute IG}
  \label{fig:ig_noise_abs}
\end{subfigure}%
\begin{subfigure}{.5\linewidth}
  \centering
  \includegraphics[width=\linewidth]{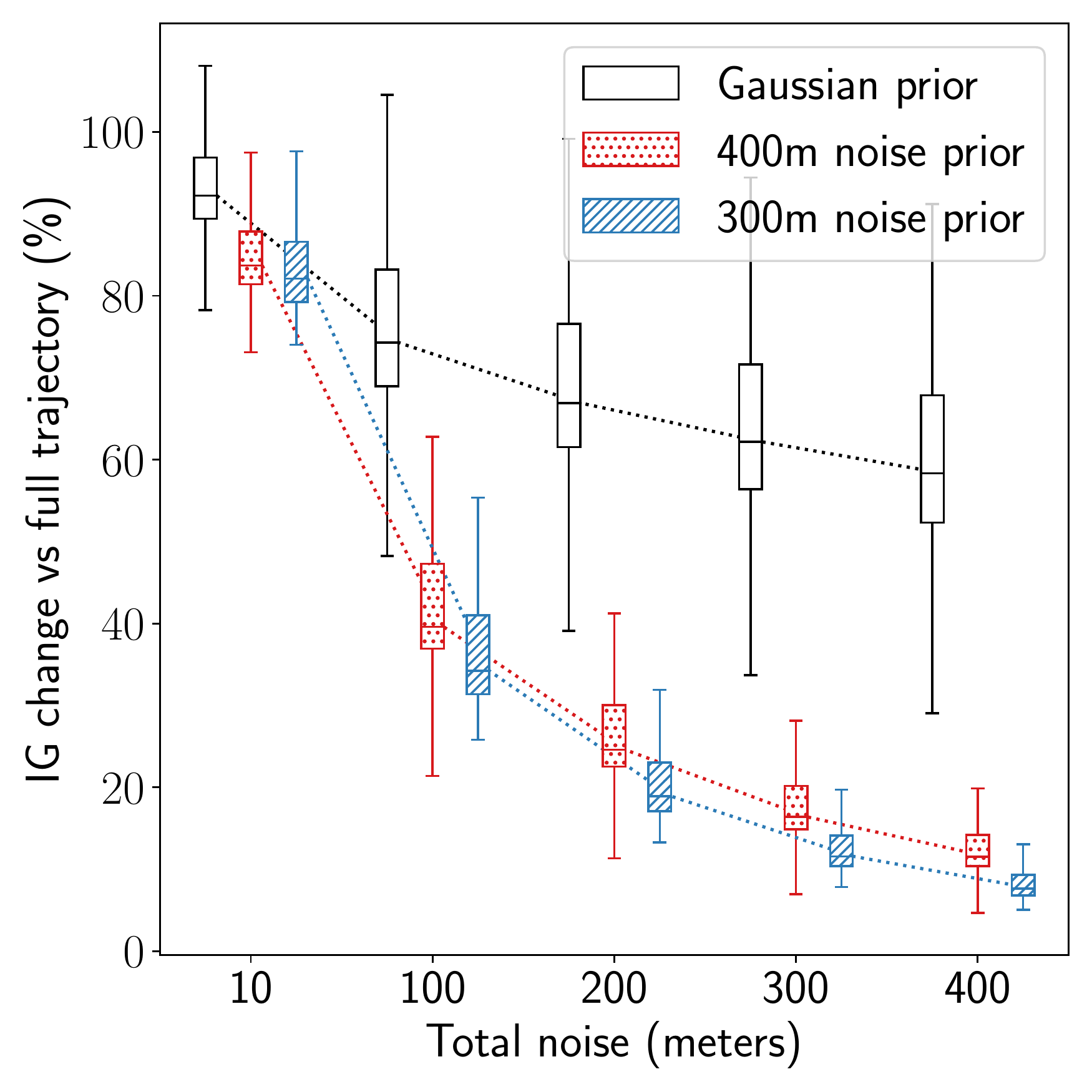}
  \caption{Percentage IG change}
  \label{fig:ig_noise_percentage}
\end{subfigure}
\caption{(\protect\subref{fig:ig_noise_abs}) Absolute IG and (\protect\subref{fig:ig_noise_percentage}) percentage IG change compared to original trajectories for different values of total noise and different types of prior knowledge.} 
\label{fig:ig_noise}
\end{figure}

We first discuss the results for perturbation. 
Figure~\ref{fig:ig_noise} shows the IG value $\ig(\trajdegraded, \priorinfo)$ and the IG ratio $\frac{\ig(\trajdegraded, \priorinfo)}{\ig(\traj, \priorinfo)}$ in percentage with different perturbations and prior knowledge. As discussed in Section~\ref{sec:exp_setup}, the total noise ranges from $3$ to $400$~meters, and the prior knowledge scenarios are \emph{Gaussion}, \emph{400m noise}, and \emph{300m noise} priors.  
Each box plot in Figure~\ref{fig:ig_noise_abs} (or~\ref{fig:ig_noise_percentage}) shows the IG value (or the IG ratio) of $\trajdegraded$ with the total noise shown on the $x$ axis given particular prior knowledge. For example, the red, dotted box plot at total noise $100$ in Figure~\ref{fig:ig_noise_abs} shows the IG values of a degraded version $\trajdegraded_{100}$ of $\traj$ with total noise $100$ meters, given that the prior knowledge is another previously-released degraded version $\trajdegraded_{400}$ of $\traj$ with total noise $400$ meters. It is clear that when total noise becomes larger, i.e., more degradation, the IG value and the IG ratio tend to decrease. 

With a more informative prior, the absolute IG value tends to be smaller. For example, the 400m noise prior is a more informative prior than the Gaussian prior, thus the absolute IG value of the same $\trajdegraded$ with the 400-noise prior is smaller than with the Gaussian prior. This also supports intuition: if one already has good information about the location of a person, getting more data does not increase such information as much as when one only has little information. This is an important property that does not exist in other baselines.

\begin{figure}[htbp!]
\centering
\begin{subfigure}{.5\linewidth}
  \centering
  \includegraphics[width=\linewidth]{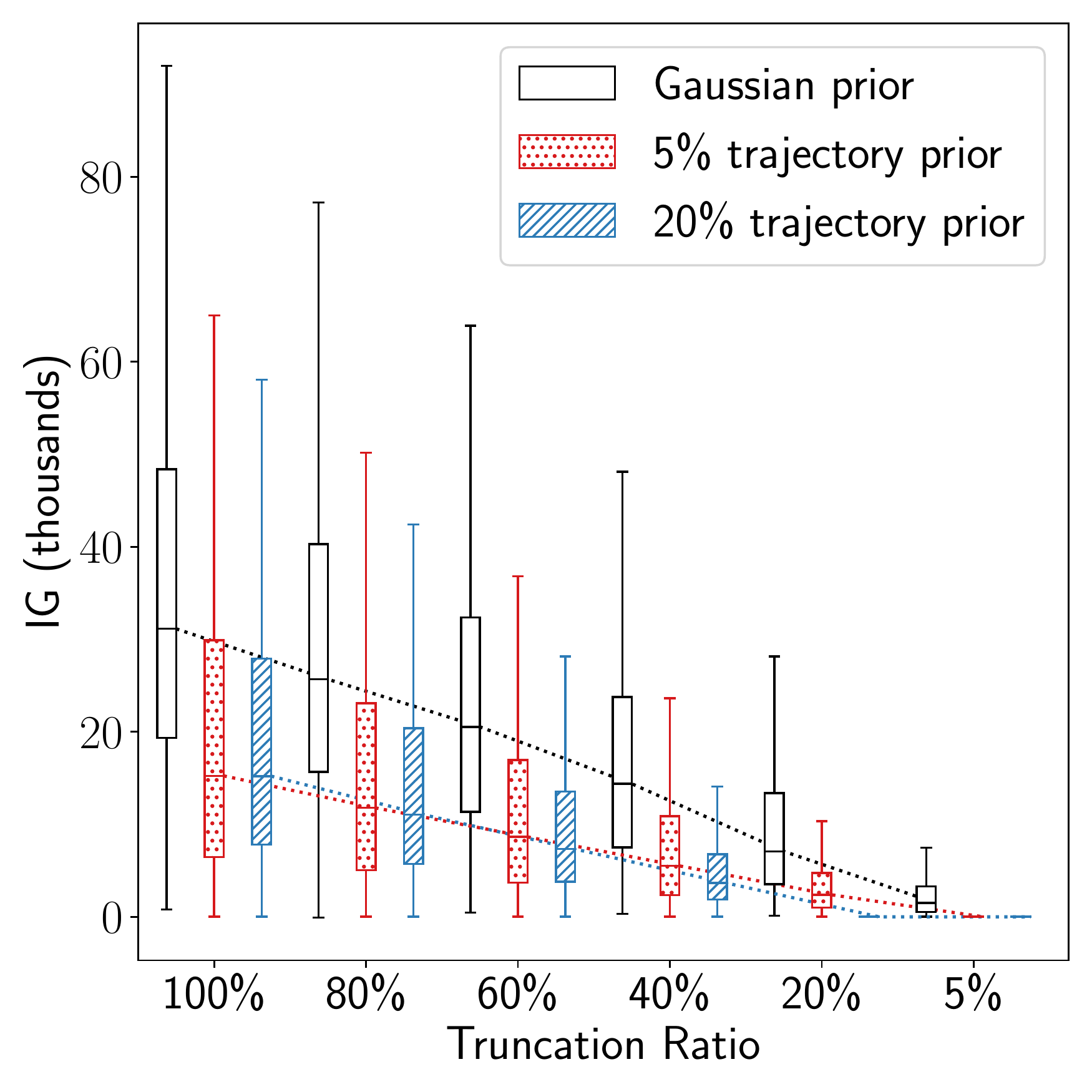}
  \caption{Absolute IG}
  \label{fig:ig_truncation_abs}
\end{subfigure}%
\begin{subfigure}{.5\linewidth}
  \centering
  \includegraphics[width=\linewidth]{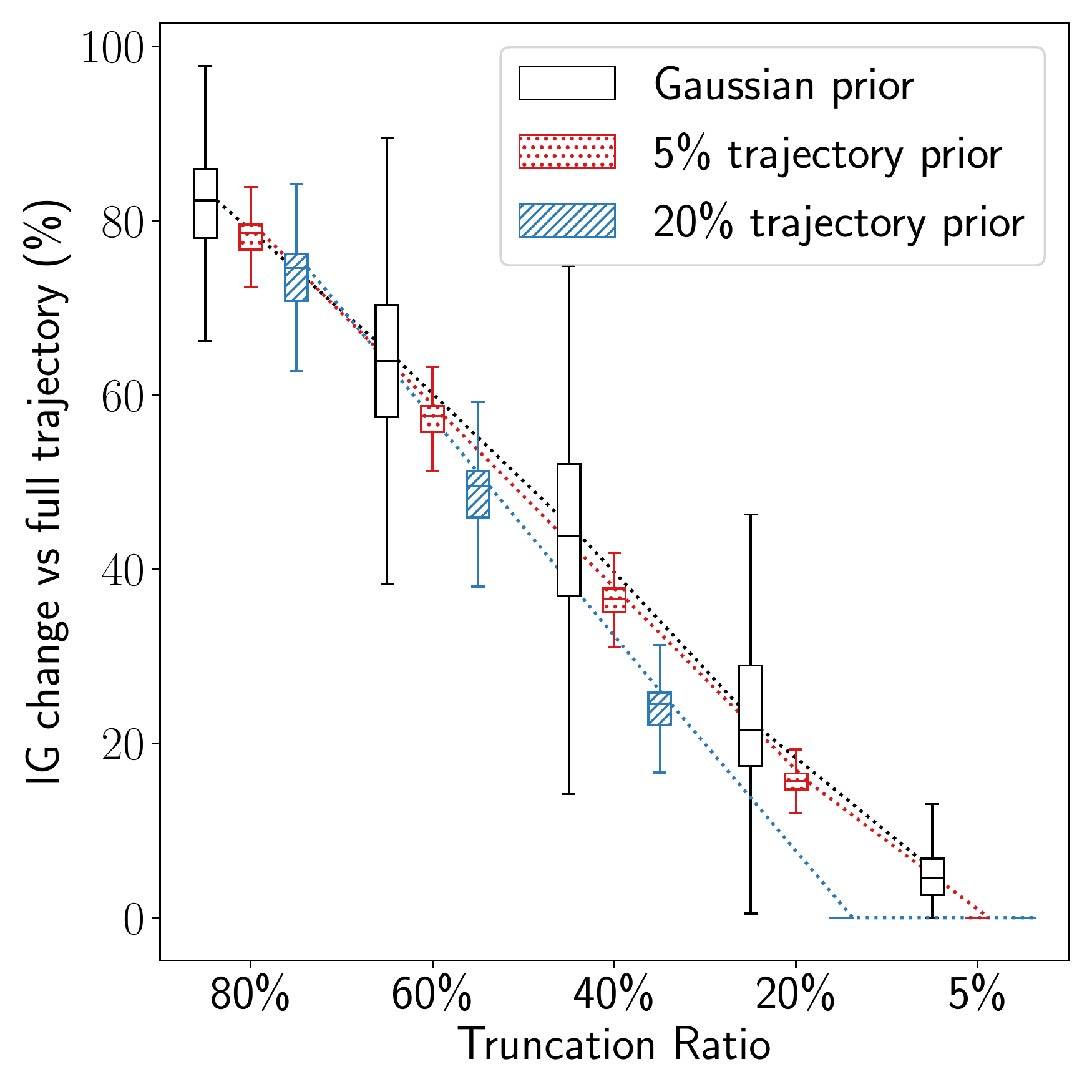}
  \caption{Percentage IG change}
  \label{fig:ig_truncation_percentage}
\end{subfigure}
\caption{(\protect\subref{fig:ig_truncation_abs}) Absolute IG and (\protect\subref{fig:ig_truncation_percentage}) percentage IG change compared to original trajectories for different truncation ratios and types of prior knowledge.}  
\label{fig:ig_truncation}
\end{figure}

\begin{figure}[htbp!]
\centering
\begin{subfigure}{.5\linewidth}
  \centering
  \includegraphics[width=\linewidth]{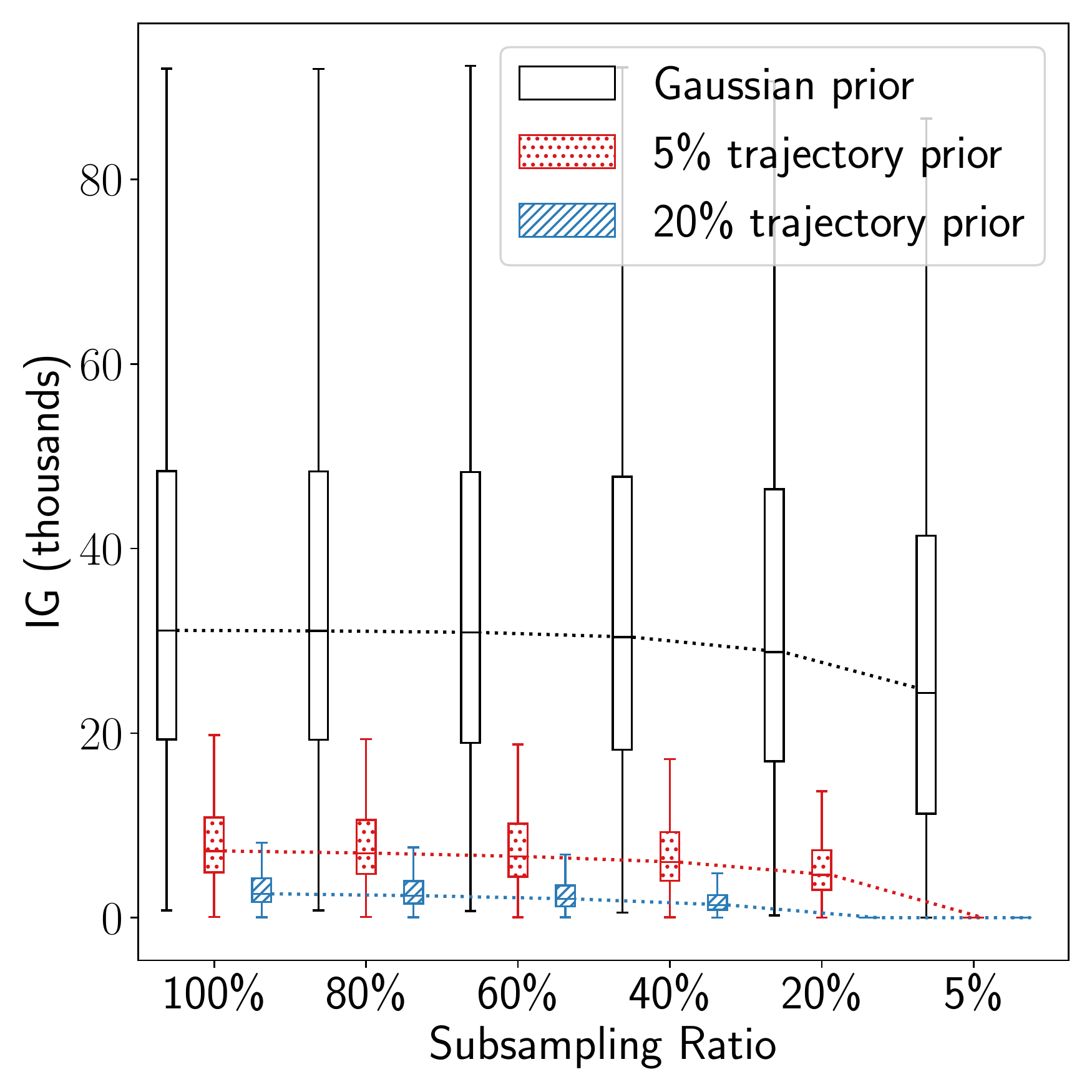}
  \caption{Absolute IG}
  \label{fig:ig_subsampling_abs}
\end{subfigure}%
\begin{subfigure}{.5\linewidth}
  \centering
  \includegraphics[width=\linewidth]{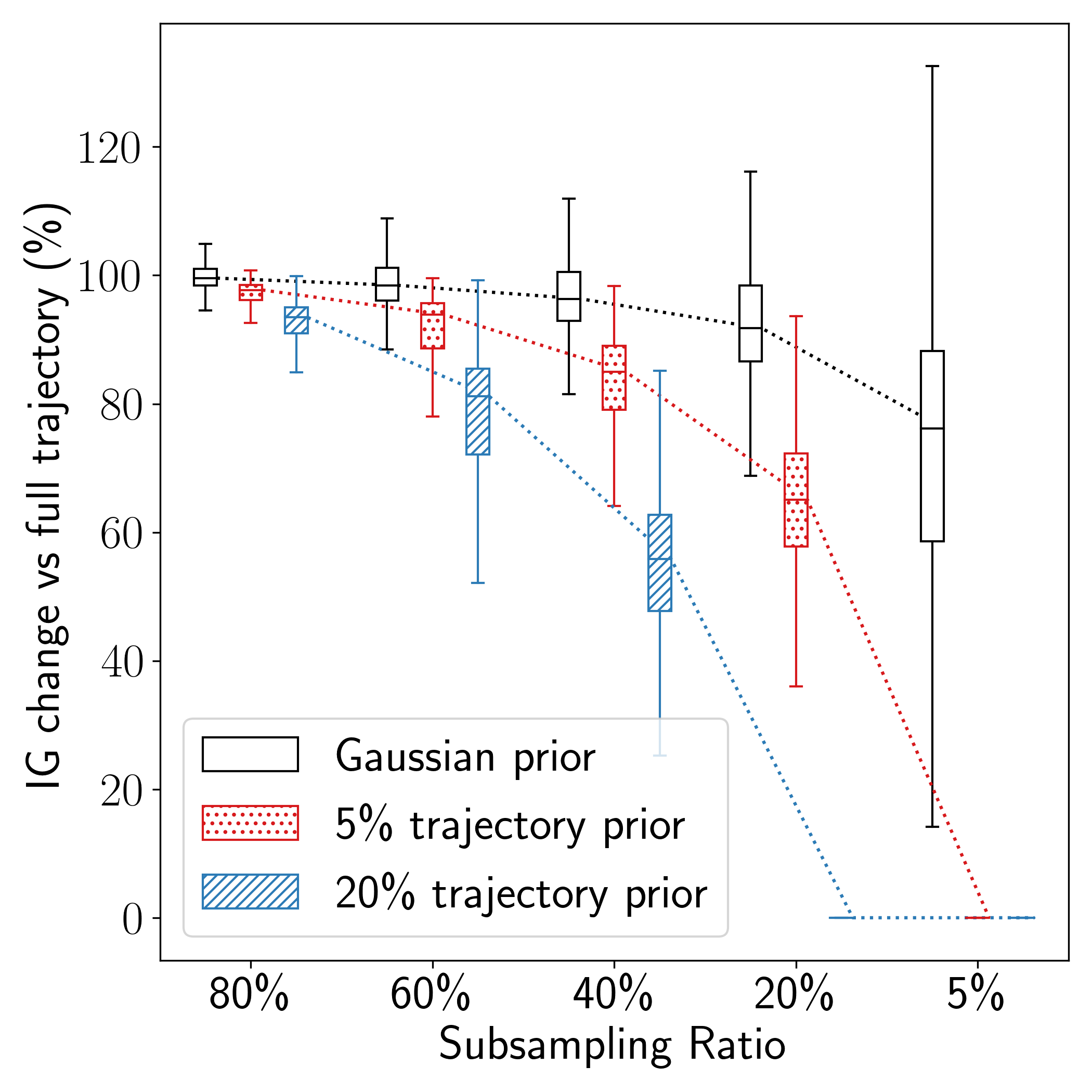}
  \caption{Percentage IG change}
  \label{fig:ig_subsampling_percentage}
\end{subfigure}
\caption{(\protect\subref{fig:ig_subsampling_abs}) Absolute IG and (\protect\subref{fig:ig_subsampling_percentage}) percentage IG change compared to original trajectories for different subsampling ratios and types of prior knowledge.}  
\label{fig:ig_subsampling}
\end{figure}

Truncation and subsampling degradation also show similar observations as those from perturbation. Figures~\ref{fig:ig_truncation} and~\ref{fig:ig_subsampling} show the IG value and the percentage of IG change for truncation and subsampling with different prior knowledge, respectively. The truncation/subsampling ratios range from $0.8$ to $0.05$, which means $80\%$ down to $5\%$ of the original trajectory is retained. Because of the $0.05$ (or $\frac{1}{20}$) ratio, these figures show the results from almost 40,000 trajectories that have at least 20 measurements. As discussed in Section~\ref{sec:exp_setup}, the prior knowledge scenarios are \emph{Gaussion prior}, \emph{$5\%$ trajectory prior} and \emph{$20\%$ trajectory prior}. The IG also tends to decrease when the trajectory is degraded more (i.e., lower ratio) and when the prior gets better (e.g, $5\%$ vs. $20\%$ priors).

The decrease in IG with truncation is much stronger than with subsampling. The reason is that knowing a measurement not only reduces the uncertainty at the time of the measurement but also for the time around that measurement. Thus when measurements are more evenly distributed, they help reduce more uncertainty, which means getting higher IG, as illustrated in Section~\ref{sec:ig_and_char_from_raw_trajectory}. 
With the same ratio, which means the same number of measurements are retained, a truncation is similar to the case when measurements are close to each other (Figure~\ref{fig:IG_qualitative_8_points_close}) while (uniform) subsampling is similar to the case when measurements are more evenly distributed (Figure~\ref{fig:IG_qualitative_8_points_dispersed}). This property also highlights the difference and advantage of IG for quantifying the VOI of a trajectory compared to other baselines.

\begin{figure}[htbp!]
\centering
\includegraphics[width=.5\linewidth]{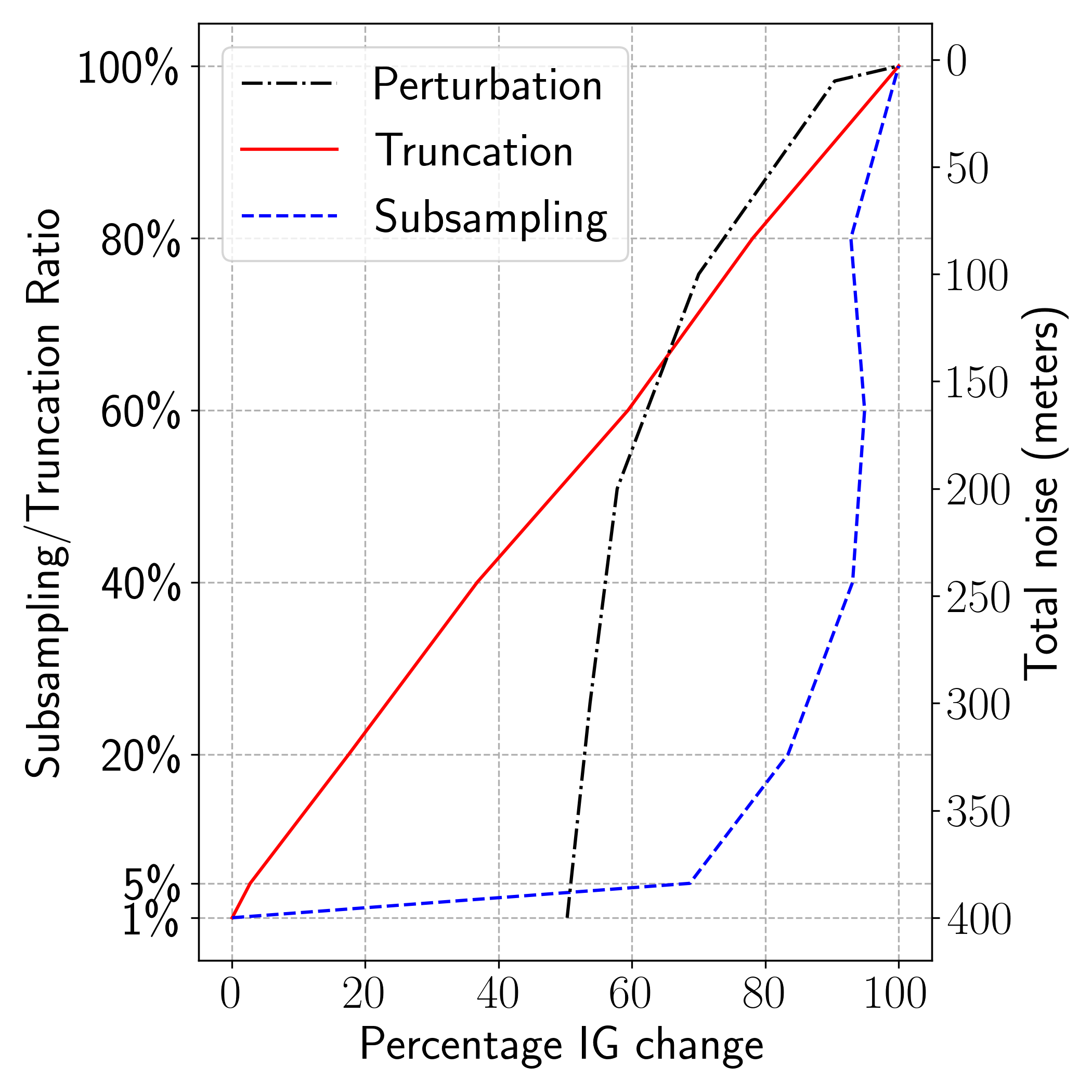}
\caption{An example of VOI equivalence classes of different types of degradation for a trajectory.}  
\label{fig:ig_degredation_equivalent}
\end{figure}

Another observation is that by effectively capturing the impact of degradation, one can find the equivalence classes of different types of degradation for the same trajectory. For instance, given $\traj$, which truncation ratio produces the same VOI compared to perturbing $\traj$ at total $300$ meter noise? Figure~\ref{fig:ig_degredation_equivalent} shows an example of such VOI equivalence for the first trajectory of the first owner in our dataset. The intersection between the (interpolated) truncation and perturbation lines indicates that truncating this trajectory at $65\%$ gives roughly similar VOI to perturbing it with total $130$m noise.

\section{Related Work}
\label{sec:related_work}
Quantifying the value of location data has been an active line of research. There have been several surveys of how individuals value their location data~\cite{cvrcek2006study, staiano2014money}. Aly et. al~\cite{aly2018value} showed how the value of a location data point can be quantified from the buyer's perspective in a geo-marketplace. Also in a geo-marketplace context, Nguyen et. al~\cite{nguyen2020spatialprivacypricing} proposed a framework allowing sellers to offer a location data point at different qualities for different prices. However, previous work focused on the monetary value of location data. To the best of our knowledge, this is the first work that attempts to quantify the intrinsic VOI of a trajectory in their most basic form, which is a sequence of location measurements. 

There is also extensive work on quantifying location privacy, which also attempted to reconstruct the locations of individuals. Krumm~\cite{krumm2009survey} surveyed a variety of computational location privacy schemes and emphasized the importance of finding a single quantifier for location privacy. Shokri et. al~\cite{shokri2011quantifying} proposed correctness as the metric for quantifying location privacy. Location privacy and the VOI of a trajectory can be related, e.g., a trajectory with high VOI may reveal more about a person than one with lower VOI, thus leaking privacy. However, location privacy does not exactly correspond to the intrinsic VOI of a trajectory: the former concerns more about semantic locations and is often more subjective, while the latter concerns more about the location coordinates over time. Correctness and subjective sensitivity, which are related to location privacy, were also discussed in detail in Section~\ref{sec:correctness}. 
\section{Conclusion and Future Work}
The intrinsic VOI of trajectories plays an important role in many applications. We proposed the IG framework and qualitatively and quantitatively demonstrated its capability to effectively capture important characteristics of raw trajectories, prior knowledge, and various types of degradation. This shows that the IG is an appropriate framework to quantify the intrinsic VOI of trajectories. There are several potential directions for future work such as incorporating other features, or supporting other degradation types.

\textbf{Acknowledgements}: This research has been funded in part by NSF grants IIS-1910950 and CNS-2027794, the USC Integrated Media Systems Center, and unrestricted cash gifts from Microsoft. Any opinions, findings, and conclusions or recommendations expressed in this material are those of the author(s) and do not necessarily reflect the views of any
of the sponsors such as the NSF.

\bibliographystyle{ACM-Reference-Format}
\scriptsize
\bibliography{bibliography}

\end{document}